\magnification=1200

\pageno=1
\centerline {\bf ON THE EXACT QUANTUM INTEGRABILITY  }
\centerline {\bf OF THE MEMBRANE    }
\medskip
\centerline {\bf Carlos Castro }
\centerline {\bf Center for Particle Theory  }
\centerline {\bf Physics Department}
\centerline {\bf University of Texas}
\centerline {\bf Austin , Texas 78757}
\centerline {\bf World Laboratory, Lausanne, Switzerland }
\smallskip

\centerline {\bf May,1996 }
\smallskip
\centerline {\bf ABSTRACT}

The  exact quantum integrability problem of the membrane is investigated. It is found that the spherical membrane moving in flat target spacetime backgrounds is an exact quantum integrable system for a particular class  of solutions of the light-cone gauge equations of motion :  a  dimensionally-reduced $SU(\infty)$ Yang-Mills theory to one temporal dimension. 
Crucial ingredients are the exact integrability property  of the $3D~SU(\infty)$ continuous Toda theory and its associated dimensionally-reduced 
$SU(\infty)$ Toda $molecule$ equation whose symmetry algebra is the $U_\infty$ algebra obtained from a dimensional-reducion of the  $W_\infty \oplus {\bar W}_\infty$ algebras that act naturally on the original $3D$ continuous  Toda theory. The $U_\infty$  algebra  is explicitly constructed in terms of exact quantum solutions of the quantized continuous Toda equation. Highest weight irreducible representations of the $W_\infty$ algebras are also studied  in detail. Continuous and  discrete energy levels are both  found in the spectrum . Other relevant topics are discussed in the conclusion.

\smallskip

\smallskip


 \centerline {\bf {1. Introduction }}
{\vbox {\vskip .5 truein}}

Recently,  exact solutions to $D=11$ spherical (super) membranes moving in flat target spacetime backgrounds were
constructed based on a particular class of  reductions of Yang-Mills equations from higher dimensions 
to four dimensions [1,2]. The starting point was dimensionally-reduced Super Yang-Mills theories based on the infinite dimensional
$SU(\infty)$ algebra. The latter algebra is isomorphic to the area-preserving diffeomorphisms of the sphere [3]. In this fashion the
super Toda molecule equation was recovered preserving one supersymmetry out of the $N=16$ expected. The expected critical target
spacetime dimensions for the ( super) membrane , $D=27 (11)$ , was closely related to that of an anomaly-free non-critical (super ) $W_{\infty}$
string theory. A BRST analysis revealed that the spectrum of the membrane should  have a relationship to the first unitary
minimal model of a $W_N$ algebra adjoined to a critical $W_N$ string in the $N\rightarrow \infty$ limit [1]. The class of particular solutions of the dimensionally-reduced $SU(\infty)$ YM equations studied by Ivanova and Popov  [2]
bears a direct relationship to $SU(\infty)$ instanton solutions in $4D$ that permits a  connection to the $SU(\infty)$ Toda molecule equation after  an specific ansatz is made [1].
 
In {\bf II} we briefly review the contents of [1] and show the crucial role that the continuous Toda equation, a $D=3$
integrable field theory, [4,5] plays  in the membrane quantization program. 
In {\bf III} an exact quantization program for  reductions of the  continuous $3D$ Toda equation is presented 
based on earlier work by Saveliev et al [5] for the finite $N$ case . Armed with these exact quantum solutions
we built in ${\bf IV}$ an isomorphism between highest weight irreducible representations 
of $W_\infty$ algebras [6,7] and the Hilbert space of states of the quantum Toda theory. This is achieved $after$ a suitable dimensional reduction of the 
$W_\infty \oplus {\bar W}_\infty$ algebra is performed. The latter algebra is coined the $U_\infty$ algebra, and it acts  on the space of solutions of the $SU(\infty)$ Toda $molecule$ equation, a $2D$ theory, whose field is  :  $\rho (r,t)$. $r=z+{\bar z}$ ( the time coordinate ) and $t$ is a variable which appears in the construction of continuum ${\bf Z}$-graded Lie algebras : ${\cal A}_\infty (su(\infty))\sim sdiff~T^2$ which is isomorphic to the Poisson bracket algebra of the torus $T^2$ [4]. Locally, the latter algebra is also isomorphic to the Poisson bracket algebra of $S^2$, but  not globally. Also discussed in ${\bf IV}$ is the detailed explicit construction of the $U_\infty$ algebra in two different realizations and a dicussion on the Casimirs. In ${\bf V}$ discrete energy states are built for periodic-like solutions. Wavefunctional solutions in terms of Bessel's functions are found.  

Brief comments about black holes [8], non-abelian Toda theories [9], 
$W$ gravity [10], 
the fractional Quantum Hall effect [11] , universal string theory [12], 
composite antisymmetric tensor field theories of the volume-preserving diffeomorphism group [13,14], etc. are presented at the  conclusion in ${\bf VI}$. 
The idea that the membrane might be completely integrable has been proposed earlier by Fairlie et al [15]. The role of the Toda molecule in membranes was also discussed by [16,17], the super Toda molecule was studied by the author ( see references in [1]) and a review on Toda theory has been given by Toda himself in [18]. The properties  of self dual extendons has been studied by Zaikov [19]
who discovered also $W_\infty$ structures in a Chern-Simmons formulation of strings.

\centerline {\bf 2. BACKGROUND : The Toda Molecule and $SU(\infty)$ Yang-Mills}

Based on the observation that the spherical membrane (excluding the zero modes ) moving in $D$ spacetime dimensions, in the light-cone gauge, is essentially
equivalent to a $D-1$ Yang-Mills theory, dimensionally reduced to one time dimension, of the $SU(\infty)$ group [21]; 
we look for solutions of the $D=10$ Yang-Mills equations (dimensionally-reduced to one temporal dimension). 
For an early  review on membranes  see Duff [20] and the recent book by Ne'eman and Eizenberg [21].

Marquard et al [22] have shown that the light-cone gauge Lorentz algebra for the bosonic membrane is anomaly free iff $D=27$. The supermembrane critical dimension was found to be  $D=11$. To this date there is still some controversy 
about whether or not  the (super )  membrane is really anomaly free in these dimensions. They may suffer from other anomalies like $3D$ reparametrization invariance anomalies or global ones. What follows next  does $not$ depend on whether or not $D=27,D=11$ are  truly the  critical dimensions. What follows is just a straightforward quantization of a very special class of  solutions of  
the dimensionally-reduced  ( to one temporal dimension)  of $SU(\infty)$ YM equations, and which can be quantized exactly  due to  their  equivalence to the exactly integrable quantum continuous Toda molecule,  ( obtained as a dimensional-reduction of the original continuous  $3D$ Toda theory to $2D$ which is where $W_\infty$ strings live ).

Let us say that we decided to start with a $D=10~SU(\infty)$ super YM theory and we were looking for a particular class of solutions to quantize  : the solutions corresponding to  dimensional reductions to one temporal dimension. We could have chosen from the start  another values for $D$ and follow similar procedures as those presented here. At the end of the quantization program that shall be developed here one can then ask the question :'' What about the presence/absence of quantum anomalies ? What are the critical dimensions, if any ? Is the theory anomaly free in any dimesion  or just in a few ? In order to answer these questions at the $end$ of the quantization program hereby presented one needs to construct the no-ghost theorem as it happened with the string $prior$ to using the powerful tools of the BRST covariant  quantization program. The string's critical dimensions, $D=26$ ($D=10$), can be obtained by several methods and one of them is based in the no-ghost theorem!
 applied to the light-cone gauge quantization. Therefore, for the rest of this work let us put aside the issue of quantum anomalies and focus on the quantization of the $D=10$ super $SU(\infty)$ YM theory dimensionally-reduced to one temporal dimension.    

In [1] we obtained solutions to the $D=10$ YM equations dimensionally reduced to one dimension. Let us focus on the bosonic sector of the theory. The supersymmetric case can be also analyzed via solutions to the supersymmetric Toda theory which has been discussed in detail in the literature [3] . The particular class of solutions  one is interested in are those  of the type analyzed by Ivanova and Popov. Given :

$$\partial_a F_{ab} +[A_a,F_{ab}] =0.~A^\alpha_a T_\alpha \rightarrow A_a( x^b; q,p).~[ A_a,A_b] \rightarrow \{ A_a,A_b \}_{q,p}.
\eqno (2.1)$$
where the $SU(\infty)$ YM potentials [16,17] are obtained by replacing Lie-algebra valued potentials ( matrices) by $c$ number functions; Lie-algebra brackets   by
Poisson brackets w.r.t the two internal coordinates associated with the sphere; and the trace by an integral w.r.t these internal coordinates.  
In eq-(1) we performed an ansatz following the results of Ivanova and Popov with  $a,b,...   =8$ being  the transverse indices to the membrane after we performed the $10=2+8$ split of the original $D=10$ YM
equations. After the dimensional reduction to one dimension is done  we found that the following $D=10$ YM potentials, ${\cal A}$, -which will be later expressed in terms of the $D=4$ YM potentials, $A_1,A_2,A_3$ ($A_0$ can be gauged to zero )-  are
one class of solutions to the original $D=10$ equations iff they admit the following relationship :

$${\cal A}_1 =p_1 A_1.~{\cal A}_5 =p_2 A_1. ~{\cal A}_3 =p_1 A_3.{\cal A}_7 =p_2 A_3. \eqno (2.2a)$$
$${\cal A}_2 =p_1 A_2.~{\cal A}_6 =p_2 A_2.~{\cal A}_0 ={\cal A}_4 ={\cal A}_8 ={\cal A}_9 =0.  \eqno (2.2b)$$

where $p_1,p_2$ are constants and  $A_1,A_2,A_3$ are functions of $x_0,q,p$ only and obey the $SU(\infty)$ Nahm's equations :

$$\epsilon_{ijk}{\partial A_k \over \partial x_0} +\{A_i,A_j \}_{q,p} =0.~i,j,k =1,2,3. \eqno (2.3)$$

Nahm's equations were obtained from reductions of $D=4$ Self Dual Yang-Mills equations to one dimension. The temporal variable $x_o
=p_1X_0+p_2X_4$ has a $correspondence $, not an identification, with the membrane's light-cone coordinate : $X^+=X^0+X^{10}$.  We refer to Ivanova and Popov and to our results in [1,2] for details.

Expanding $A_y =\sum A_{yl}(x_o)Y_{l,+1}.~A_{\bar y} =\sum A_{{\bar y}l}(x_o)Y_{l,-1}$; and $A_3$ in terms of $Y_{l0}$,
the ansatz which allows to recast the $SU(\infty)$ Nahm's equations as a Toda molecule equation is [1] :

$$\{A_y,A_{\bar y} \} =-i\sum^{\infty}_{l=1}~exp[K_{ll'}\theta_{l'}]Y_{l0} (\sigma_1,\sigma_2).~A_3
=-\sum^{\infty}_{l=1}{\partial \theta_l\over \partial \tau}.Y_{l0}. \eqno (2.4)$$ 

with $A_y ={A_1+iA_2\over \sqrt 2}.~A_{\bar y} ={A_1-iA_2\over \sqrt 2}.$

Hence, Nahm's equations become :

$$-{\partial^2 \theta_l\over \partial \tau^2 } =e^{K_{ll'} \theta_{l'}}.~~ l,l'=1,2,3....\eqno (2.5)$$

This is the $SU(N)$ Toda molecule equation in Minkowski form. The $\theta_l$ are the Toda fields where $SU(2)$ has been embedded
minimally into $SU(N)$. $K_{ll'}$ is the Cartan matrix which in the continuum limit becomes :$\delta''(t-t')$ [2,3]. The solution of the Toda theory is well known to the experts by now.  Solving for the  $\theta_l (\tau)$ fields and plugging their values into the first term of eq-(2.4) yields an infinite number of equations -in the $N\rightarrow \infty$ limit- for the infinite number of ``coefficients'' $A_{yl} (x_o),A_{{\bar y}l}(x_o)$. This allows to solve for the YM potentials $exactly$.  The ansatz [1] automatically yields the coefficients in the expansion of the $A_3$ component of the $SU(\infty)$ YM field given in the second term of (2.4). Upon quantization of the $SU(\infty)$ YM theory, the first term in eq-(2.4) is replaced by a commutator of two operators and as such the coefficients  $A_{yl} (x_o),A_{{\bar y}l}(x_o)$ become operators as well. The Toda fields become also operators in the Heisenberg representation. We will discuss in detail the quantization of the Toda!
 theory by taking the continuum limit of [5]. The important point is that an exact  quantization of the integrable Toda theory yields a quantization of the $SU(\infty)$ YM theory and hence it furnishes an exact quantum integrable system.   

The
continuum limit of (2.5) is 

$$-{\partial^2 \theta (\tau,t) \over \partial \tau^2 } =exp~[\int~dt'\delta''(t-t') \theta (\tau,t')]. \eqno (2.6)$$

Or in alternative form :

$$-{\partial^2 \Psi(\tau,t) \over \partial \tau^2 } = \int~\delta'' (t-t')exp[\Psi (\tau,t')]~dt' ={\partial^2 e^{\Psi }\over
\partial t^2}. \eqno (2.7)$$

if one sets $K_{ll'}\theta_{l'} =\Psi_l$. The last two equations are the dimensional reduction of the $3D\rightarrow 2D$
continuous Toda equation given by Leznov and Saveliev :

$${\partial^2 u(\tau,t)\over \partial \tau^2} =-{\partial^2 e^u\over \partial t^2}.~i\tau \equiv r=z+{\bar z}. \eqno (2.8a)$$
Eq-(2.8a) is referred as the $SU(\infty)$ Toda $molecule$ whereas 

$${\partial^2 u(z,{\bar z},t)\over \partial z \partial {\bar z}} =-{\partial^2 e^u\over \partial t^2}.\eqno (2.8b)$$
is the $3D$ continuous Toda equation which can 
obtained as rotational Killing symmetry reductions of Plebanski equations for Self-Dual Gravity in
$D=4$. Eq-(2.8a) is an effective $2D$ equation and in this fashion the original $3D$ membrane can be related to a $2D$ theory ( where the $W_\infty$ string lives in )  after the light-cone gauge is chosen.

In [1] we established the correspondence between the target space-times of non-critical $W_{\infty}$ strings and that of
membranes in $D=27$ dimensions. The supersymmetric case was also discussed and $D=11$ was retrieved. The skeptic reader who is not convinced that such a correspondence is possible may jump ahead directly to section {\bf III}. As we emphasized earlier :  whether or not such a correspondence between non-critical $W_{\infty}$ strings and membranes is possible ; whether or not $D=27,D=11$ are truly the critical dimensions;.... the results that follow from {\bf III} onwards do $not$ depend and do $not$ rely in 
anyway whatsoever in the remainder  of this section !. The reason we believe the rest of this section is relevant  will become clear  when we reach the end of the road and ask again :'' What about the anomalies and the construction of a physical spectrum ?''. In order to answer these  questions  we must  have a list of all unitary highest weight irreducible representations for $W_{\infty}$ algebras and develop afterwards  the no-ghost theorem. This will require another series of articles.   

The relevance of developing a $W_{\infty}$ conformal field theory has been emerging over the past years [43]. It was shown in [23] that
the effective induced action of $W_N$ gravity in the conformal gauge takes the form of a Toda action for the scalar fields and
the $W_N$ currents take the familiar free field form. The same action can be obtained from a constrained $WZNW$ model. Each of
these Toda actions posseses a $W_N$ symmetry. The authors [23,24] coupled $W_N$ matter to $W_N$ gravity in the conformal gauge,
and integrating out the matter fields, they arrived at the induced effective action which was precisely the same as the Toda
action. Non-critical $W_N$ strings are constructed the same way. The matter and Liouville sector of the $W_N$ algebra can be
realized in terms of $N-1$ scalars, $\phi_k,\sigma_k$ repectively. These realizations in general have background charges which
are fixed by the Miura transformations [25,26]. The non-critical string is characterized by the central charges of the matter and
Liouville sectors, $c_m,c_L$. To achieve a nilpotent BRST operator these central charges must satisfy :

$$c_m+c_L =-c_{gh} =2\sum^N_{s=2} (6s^2-6s+1) =2(N-1)(2N^2+2N+1). \eqno (2.9)$$

In the $N\rightarrow \infty$ limit a zeta function regularization yields $c_m+c_L =-2$.

We were able to show that the critical membrane background in $D=27$ was the same as that of a non-critical $W_{\infty}$ string 
background if one adjoined the first unitary minimal model of the $W_N$ algebra to that of a critical $W_N$ string spectrum in
the $N\rightarrow \infty$ limit. In particular :

$$c_{eff} =D=1-12x^2 =(1-12x^2_o) +c_{m_o} = 26 -(1-{6\over (N+1)(N+2)}) +2{N-1\over N+2} \Rightarrow D=27. \eqno (2.10)$$
similar arguments led us to $D=11$ in the supermembrane case [1].

The value for the total central charge of the matter sector is $c_m =2+{1\over 24}$ after a zeta function regularization.
That of the Liouville sector is $c_L =-4-{1\over 24}$. These values bear an important connection to the notion of Unifying $W$
algebras [27]. It happens that when the central charges have for values at $c(n)={2(n-1)\over n+2};{2(1-2n)\over
n-2};-1-3n$, there exists a Unifying Quantum Casimir $W$ algebra :  

$${\cal W}{\cal A}_{n-1} \leftrightarrow {\cal W} (2,3,4,5) \sim {sl {\hat (2,R)}_n \over {\hat U(1)}}. \eqno (2.11)$$ 

in the sense that these algebras truncate at degenerate values of the central charge to a smaller algebra.

We see that the value $c_m$ after regularisation corresponds to the central charge of the first unitary minimal model of
$WA_{n-1}$ after $n$ is analytically continued to a negative value $n=-146 \Rightarrow 2(n-1)/(n+2) =2+1/24.$
The value of $c_L$ does not correspond to a minimal model but it also corresponds to a special value of $c$ where the
$WA_{n-1}$ algebra truncates to that of the unifying coset : $c(n)= 2(1-2n)/(n-2) =-4 -1/24$ for $n=146$.
In virtue of the
quantum equivalence equivalence between negative rank $A_{n-1}$ Lie algebras with $n=146\rightarrow -n=-146; c=2+{1\over 24}$
and  ${\cal W} (2,3,4,5)\sim W_{\infty}$ at $c=2$; i.e. the Hilbert spaces are isomorphic , 
we can study the spectrum of non-critical ${\cal W} (2,3,4,5)$ strings and claim that it ought to give very relevant
information concerning the membrane's spectrum.

Unfortunately, non-critical ${\cal W} (2,3,4,5)$ strings are prohibitively 
complicated. One just needs to look into the cohomology of ordinary critical $W_{2,s}$ strings to realize this [25].
Nevertheless there is a way in which one can circumvent this problem. The answer lies in the integrabilty property of the
continuous Toda equation [4,5] and the recently constructed quasi-finite highest weight irreducible representations of
$W_{1+\infty}, W_{\infty}$ algebras [6,7]. For this purpose one needs to compute explicitly the value of the coupling constant
appearing in the exponential potential of the Quantum Toda theory [1]. The latter is conformally invariant and the conformally
improved stress energy tensor obeys a Virasoro algebra with an adjustable central charge whose value depends on the coupling
constant $\beta$ appearing in the potential [15]. This value coincides precisely with the one obtained from a Quantum
Drinfeld-Sokolov reduction of the $SL(N,R)$ Kac-Moody algebra at level $k$ :

$$\beta ={1\over \sqrt {k+N}}.~~~c(\beta) =(N-1) -12 |\beta \rho -(1/\beta) {\tilde \rho})|^2. \eqno (2.12)$$

where $\rho,{\tilde \rho}$ are the Weyl vectors of the (dual) $A_N$ algebra. We see [1,23] that one can now relate the value of
the background charge $x$ in (10)  and $\beta$ when $k=-\infty.~N=\infty.~k+N=constant$ :

$$2x^2=(-13/3) =({1\over \sqrt {k+N}} -\sqrt {k+N})^2 = (\beta -{1\over \beta} )^2 \Rightarrow \beta^2 = {-7+(-)\sqrt 13 \over
6}.    .\eqno (2.13)$$

so $\beta$ is purely imaginary. This should not concern us.
There exist integrable field theories known as Affine Toda theories whose coupling is imaginary but posseses soliton solutions
with real energy and momentum [28].

This completes the review of [1]. We hope that we've clarified the interplay between $3D$ and $2D$ that appears after the light-cone gauge for the membrane is chosen and the importance of eqs-(2.8a,2.8b) after the ansatz (2.4) is made. We assume that such a correspondence holds off-shell as well. 

\bigskip
\centerline {\bf { 3 }}
\centerline{\bf The Quantization of the Continuous Toda Field}
{\vbox {\vskip .2 truein}}   

Having reviewed the essential results of [1] permits us to look for classical solutions to the 
continous Toda equation and to implement the Quantization program presented in [5] after one takes the continuum limit which is what we are going to do in this section. 
The general solution to (2.8a) depending on two
variables , say $r\equiv z_+ +z_-$ and $t$ (not to be confused with time ) was given by [4]. 
The solution is determined by two
arbitrary functions , $\varphi (t)$ and $d(t)$. It is :

$$exp[-x(r,t)]=exp[-x_o(r,t)]\{1+\sum_{n> 1}~(-1)^n\sum_\omega\int
\int....exp[r\sum_{m=1}^n~\varphi (t_m)]~ \prod^{m=n}_{m=1}~dt_m d(t_m)$$
$$[\sum_{p=m}^n~\varphi
(t_p)]^{-1}[\sum_{q=m}^n~\varphi (t_{\omega (q)})]^{-1}. [\epsilon_m (\omega)\delta
(t-t_m)-\sum_{l=1}^{m-1}~\delta''(t_l-t_m)\theta [\omega^{-1}(m)-\omega^{-1}(l)]]\}. \eqno (3.1)$$
with :
$\rho_o=\partial^2x_o/\partial t^2 =r\varphi (t) +ln~d(t).$ This defines the boundary values of the solution 
$x(r,t)$ in the asymptotic region $r\rightarrow \infty $.
$\theta$ is the Heaviside step-function. 
$\omega$ is any permutation of the indices from $[2..........n] \rightarrow [j_2,..........j_n]$. 
$\omega (1)\equiv 1.$ $\epsilon_m (\omega)$ is a numerical coefficient. See [4] for details. 

An expansion of (3.1) yields :
$$exp[-x]=exp[-x_o]\{1-\mu +{1\over 2}\mu^2+........\}.\eqno (3.2)$$
where :
$$\mu \equiv {d(t)exp[r\varphi (t)]\over \varphi^2}. \eqno  (3.3)$$

The solution to the Quantum $A_{\infty}$  ( continous) Toda chain can be obtained by taking the continuum limit of 
the general solution to the finite nonperiodic Toda chain associated with the Lie algebra 
$A_N$ in the $N\rightarrow \infty$ limit. This is performed by taking the continuum limit of eqs-(30-34) and  eqs-(82-86) of [5] :

$$\varphi_i \rightarrow x_o(r,t).~\psi_{j_s} \rightarrow \partial^2 x_o/\partial t^2_s =r\varphi (t_s) +ln~d(t_s). \eqno (3.4)$$
In the $r\rightarrow \infty$ limit the latter tends to $r\varphi (t_s)$.
The continuum  limit of (86) in [5] is :

$$\sum_{j_1j_2...j_n} \rightarrow \int~\int~....dt_1 dt_2.......dt_n.~{\cal P}^1 \rightarrow [\sum \varphi (t_p) +O(\hbar )]^{-1}.
{\cal P}^2 \rightarrow [\sum \varphi (t_{\omega (q)}) +O(\hbar )]^{-1}.
\eqno (3.5)$$

Therefore, one just has to write down the quantum corrections 
to the  two factors $[\sum~\varphi]^{-1}$ appearing in  eq-(3.1) above. 
One must replace the first factor by a summation from $p=m$ to $p=n$ of terms like :
$$[\varphi (t_p) +O(\hbar)] \rightarrow \varphi (t_p)-{i \hbar \over w(t_p)} [{1\over w (t_p)}]_{,t_pt_p} -i \hbar\sum^n_{l=p+1} 
 {1\over w(t_l)} [{1\over w (t_l)}]_{,t_lt_l}. \eqno (3.6)$$ 

and the second factor  by  a summation from $q=m$ to $q=n$ of terms like :

$$[\varphi (t_{\omega (q)}) +O(\hbar )] \rightarrow [eq~(3.6) : ~p\rightarrow \omega (q)] +i \hbar \delta (t-t_{\omega (q)})$$
$$-i\hbar \sum_{l=1}^{q-1} [ w(t_{\omega (l)})]^{-1} [{1\over w(t_{\omega (l)})}]_{,t_{\omega (l)} t_{\omega (l)}} 
+i \hbar
\sum_{l=q+1}^n  [ w(t_{\omega (l)})]^{-1}     [{1\over w(t_{\omega (l)})}]_{,t_{\omega (l)} t_{\omega (l)}}    \eqno (3.7)$$

where $w (t)$ is a positive function that is the continuum limit of eqs-(30,34) of [5]. What one has done is to replace :

$${\hat k}_{j_mj_l}\equiv {{\tilde  k}_{j_mj_l}\over w_{j_l} w_{j_m}} \rightarrow \int~dt_m {\delta'' (t_m-t_l)\over w(t_l) w (t_m)} =
  {1\over w(t_l) }[{1\over w (t_l)}]_{,t_lt_l}\eqno (3.8a)$$
in all the equations in the continuum limit. One may  smear out the delta functions which appears in eq-(3.7) if one wishes so that  the denominators of eq- (3.1) are well defined.

These are the quantum corrections to the classical solution $\rho = \partial^2 x/\partial t^2 $ where $x(r,t )$ is given
in (3.1). These  are the continuum limits of eqs-(82-86) of [5].  It is important to realize that one must not add quantum
corrections to the $\varphi, d(t) $ appearing in the terms $exp[r\sum \varphi]$ and $x_o$ of (3.1). The former are two arbitrary
functions which parametrize the space of classical solutions. Upon quantization it follows from eq-(31) of [5] that $\varphi (t),d(t)$ become $r$ independent
( ``time'' independent ) operators obeying the equal $r$ ( time) commutation relations given by eq-(33) of [5] in the continuum limit  :
$$[\varphi (t),ln~d(t)]=-i{\hbar}{1\over w(t) }[{1\over w (t)}]_{,tt}.
\eqno (3.8b)$$
Therefore,  $\rho$ and $x$ acquire ${\hbar}$ quantum corrections given by 
(3.6,3.7) through the $c$-number function $w(t)$ terms and depend on  the non-commuting operators given by $\varphi (t), ln~d(t)$. Hence, $\rho$ or  $x$ should be seen as quantum operators acting on the Hilbert space of states associated with the quantization of the continuous Toda field : $\rho (r,t)$. Such states are
always  labeled as $|\rho>_{\varphi (t),d(t)}$. For convenience purposes we shall omit the suffix from now on but we should keep ithis in mind. 
Upon quantization, ${\hbar}$ appears and associated with Planck's constant a new parametric function has to appear : $w(t)$. One
has to incorporate also the coupling constant $\beta$ in all of the equations . This is achieved by rescaling the continuous
Cartan matrix by a factor of $\beta$ so that  $\partial^2 x/\partial t^2$ and  $\partial^2 x_o/\partial t^2$ are rescaled  by a
factor of $\beta$; i.e. $r\varphi$ acquires a factor of $\beta$ and $d(t)\rightarrow d(t)^\beta$.  
 Since $\beta$ is pure imaginary, for convergence purposes in the
$r=\infty$ region we must have that $\beta \varphi <0\Rightarrow \varphi =i\varphi$ also. In the rest of this section we will
work without the $\beta$ factors and only reinsert them at the end of the calculations. There is nothing unphysical about this
value of $\beta$ as we said earlier.

One of the integrals of motion is the energy. The continuos Toda chain is an exact integrable system in the sense 
that it posesses an infinite number of functionally independent integrals of motion : $I_n (p,\rho)$ in involution. i.e. 
The Poisson brackets  amongst $I_n,I_m$ is zero. Since these are integrals of motion, they do not depend on $r$. These
integrals can be evaluated most easily in the asymptotic region $r\rightarrow \infty$. This was performed in [2] for the case
that  $\varphi (t)$ was a negative real valued function which simplified the calculations. 
For this reason the energy
eigenvalue given in [2] must now be  rescaled by a factor of $\beta^2$  :  

$$E=\beta^2\int^{2\pi}_0~dt (\int^t~dt'\varphi (t'))^2. \eqno (3.9 
)$$

where we have chosen the range of the $t$ integration to be $[0,2\pi]$. Since $\beta\varphi <0 \Rightarrow \beta^2\varphi^2 >0$
and the energy is positive. We insist, once more, that $t$ is a parameter which is not the physical time and that $\varphi (t)$
does not acquire quantum corrections. The latter integral (22) is the eigenvalue of the Hamiltonian which is one of the Casimir
operators for the irreducible representations of $A_N$ in the $N\rightarrow \infty$ limit.  

The authors

[]

found discrete energy levels for the quantum mechanical $SU(N)$ YM model. They did emphasize that discontinuities can occur in the $N=\infty$ limit. The reason that for finite $N$ a discrete spectrum occurs is due to the existence of a non-zero value of the Casimir energy. The classical membrane admits continuous deformations of zero area with no energy expense . Quantum mechanically, this classical instability is cured by quantum effects and any wave-function gets stuck in the potential valleys that become increasingly narrow the farther out one gets. Finite-energy wave functions fall-off rapidly and the quantum Hamiltonian is purely discrete. The ground state energy corresponds to the finite non-zero Casimir energy : the point beyond which no further deformations of the membrane into long stringlike configurations is possible.
In the supermembrane case the Casimir energy is zero and a continuous spectrum emerges. However, this was performed at a $finite$ value of $N$. Here we are 
discussing the different case ;  what happens at the $N=\infty$ limit.

\medskip
\centerline{\bf 4}
\centerline{\bf 4.1.  Highest Weight Representations}

We can borrow now the results by [6,7] on the quasi-finite highest weight irreducible representations of $W_{1+\infty} $
and $W_{\infty}$ algebras. The latter is a subalgebra of the former.  For each highest weight state,$|\lambda >$ parametrized by
a complex number $\lambda$ the authors [6,7] constructed  representations consisting of a finite number of states at each
energy level by succesive application of ladder-like operators. A suitable differential constraint on the generating function
$\Delta (x)$ for the highest weights $\Delta^\lambda_k$ of the representations was necessary in order to ensure that, indeed,
one has a finite number of states at each level. The highest weight states are defined :

$$W(z^n D^k) |\lambda> =0.~n\ge 1.k\ge 0.~~W(D^k) |\lambda>=\Delta^\lambda_k |\lambda > .~k\ge 0. \eqno (4.1)$$

The $W_{1+\infty}$ algebras can be defined as central extensions of the Lie algebra of differential operators on the circle.
$D\equiv zd/dz.~n\epsilon {\cal Z}$ and $k$ is a positive integer. The generators of the $W_{1+\infty}$ algebra are denoted by 
$W(z^n D^k)$ and the $W_{\infty} $ generators  are obtained from the former : ${\tilde W} (z^n D^k) =W (z^n D^{k+1})$; where 
$(n,k \epsilon {\cal Z}.~k\ge 0)$. (There is no spin one current).
The generating function $\Delta (x)$ for the weights is :

$$\Delta (x) = \sum_{k=0}^{k=\infty} \Delta^\lambda_k~{x^k\over k!}. \eqno (4.2)$$
where we denoted explicitly the $\lambda$ dependence as a reminder that we are referring to the highest weight state $|\lambda
>$ and satisfies the differential equation required for quasi-finiteness :

$$b(d/dx)[(e^x-1)\Delta (x) +C] =0.~ b(w) =\Pi~(w-\lambda_i)^{m_i}.~\lambda_i\not= \lambda_j. \eqno (4.3)$$

$b(w)$ is the characteristic polynomial. $C$ is the central charge and the solution is :

$$\Delta (x) ={\sum_{i=1}^K~p_i(x)e^{\lambda_i x} -C\over e^x-1}.   \eqno (4.4)$$
The generating function for the $W_{\infty}$ case is ${\tilde \Delta} (x) =(d/dx) \Delta (x)$ and the central charge is 
$c=-2C$.

The Verma module is spanned by the states :

$$|v_\lambda> =W(z^{-n_1}D^{k_1})W(z^{-n_2}D^{k_2})...........W(z^{-n_m}D^{k_m})|\lambda >. \eqno (4.5)$$

The energy level is $\sum_{i=1}^{i=m}~n_i$. For further details we refer to 
[6,7]. Highest weight unitary representations for 
the $W_{\infty}$ algebra obtained from field realizations with central charge $c=2$ were constructed in [6].

The weights associated with the highest weight state $|\lambda >$ will be obtained from the expansion in (4.2).
In particular, the "energy" operator acting on $|\lambda >$ will be  :

$$W(D)|\lambda > =\Delta^\lambda_1 |\lambda >. \eqno (4.6)$$ 
$L_o = -W(D)$ counts the energy level :$[L_o, W(z^n D^k)] =-nW(z^n D^k)$.

As an example we can use for $\Delta (x)$ the one obtained in the free-field realization by free fermions or {\bf bc} ghosts 
[6]

$$ \Delta (x) =C{e^{\lambda x}-1\over e^x -1} \Rightarrow \partial \Delta/\partial \lambda =C{xe^{\lambda x}\over e^x -1}.
\eqno (4.7)$$
The second term is the generating function for the Bernoulli polynomials :

$${xe^{\lambda x}\over e^x -1} = 1+(\lambda -1/2)x +(\lambda^2-\lambda +1/6){x^2\over 2!} + (\lambda^3 -3/2
\lambda^2+1/2\lambda){x^3\over 3!} +.........\eqno (4.8)$$

Integrating (4.8) with respect to $\lambda $ yields back :

$$\Delta (x) =C{e^{\lambda x}-1\over e^x -1} =\sum_{k=0}~\Delta_k{x^k\over k!}. \eqno (4.9)$$

The first few weights (modulo a factor of $C$) are then :

$$\Delta_0= \lambda.~\Delta_1 = (1/2) (\lambda^2 -\lambda).~\Delta_2 = (1/3)\lambda^3 -(1/2)\lambda^2 +(1/6) \lambda.....\eqno
(4.10)$$

The generating function for the $W_{\infty}$ case is ${\tilde \Delta} (x) ={d\Delta (x)\over dx}\Rightarrow {\tilde
\Delta}^\lambda_k =\Delta^\lambda_{k+1}$.
This completes the short review of the results in [6,7]. Now we proceed to relate the construction of [6,7] with the results of section {\bf III}. 
\smallskip
\centerline{\bf 4.2  The $U_\infty$ Algebra}
\smallskip
We are going to construct explicitly the dimensional reduction of the 
$W_\infty \oplus {\bar W}_\infty $ algebra, the $U_\infty$ algebra, in terms of what one knows from the previous results in ${\bf 4.1}$.
>From the previuos discussion we learnt that 
${\tilde \Delta}^\lambda_1=\Delta^\lambda_2$ is the weight associated with the "energy" operator. In the ordinary string, $W_2$
algebra, the Hamiltonian is related to the Virasoro generator, $H=L_o+{\bar L}_o$ and states are built in by applying the ladder-like
operators to the highest weight state, the "vacuum". In the $W_{1+\infty},W_{\infty}$ case it is $not$ longer true, as we shall see,  that the
Hamiltonian ( to be given later ) can be written exactly in terms of the zero modes w.r.t the $z,{\bar z}$ variables of the $W_2$ generator, once the realization of the
$W_{\infty}$ algebra is given  in terms of the $dressed $ continuous Toda field , $\Theta (z,{\bar z},t)$,  given by Savaliev [4] :

$$ {\tilde W}^+_2  =\int^{t_o}~dt_1~\int^{t_1}~dt_2~exp[-\Theta (z,{\bar z};t_1)]{\partial\over \partial
z}exp[\Theta (z,{\bar z};t_1)-\Theta (z,{\bar z};t_2)] {\partial\over \partial z} exp [\Theta (z,{\bar z};t_2)]. \eqno (4.11)$$
>From now on we shall omit the label dressed continuous Toda field in our equations. The expression relating the dressed field in terms of the ordinary solutions of the continuous Toda equation is given in [2] and shall not be repeated here.   
The chiral generator has the form $W^+_{h,0} [\partial \rho/\partial z....\partial^h \rho /\partial z^h]$ [2]  
and the similar expression for the antichiral generator $ W^-_{0,{\bar h}}$ is obtained by replacing $\partial_z\rightarrow 
\partial_{{\bar z}}$ in (4.11). 
After a dimensional reduction from $D=3\rightarrow D=2$ is taken, $r=z+{\bar z}$, one has :

$${\tilde W}_2 (r,t_o) =\int^{t_o}~dt_1~\int^{t_1}~dt_2~exp[-\rho (r,t_1)]{\partial\over \partial r}exp[\rho (r,t_1)-\rho
(r,t_2)] {\partial\over \partial r} exp [\rho (r,t_2)]. \eqno (4.12)$$

When $\rho
(r,t)$ is quantized in eq-(3.6,3.7) it becomes an operator, ${\hat \rho}(r,t)$, acting on a suitable Hilbert space of states, say
$|\rho>$, and in order to evaluate (4.12) one needs to perform the highly complicated Operator Product Expansion between the
operators ${\hat \rho} (r,t_1),{\hat \rho} (r,t_2)$. Since these are no longer free fields it is no longer trivial to compute
per example the operator products :

$${\partial \rho \over \partial r}.e^\rho.~~e^{\rho (r,t_1)}.e^{\rho(r,t_2)}.....\eqno (4.13)$$

Quantization deforms the classical
$w_{\infty}$ algebra into $W_{\infty}$ [29,30]. For a proof that the $W_{\infty}$ algebra is the Moyal bracket deformation of
the $w_{\infty}$ see [30]. Later  in [31] we were able to construct the non-linear ${\hat W}_{\infty}$ algebras from
non-linear integrable deformations of Self Dual Gravity in $D=4$. 
Since the $w_{\infty}$ algebra has been effectively quantized the
expectation value of the ${\tilde W}_2$ operator in the $\hbar \rightarrow 0$ limit, is related to the ${\tilde W}_2
(classical)$  given by (4.12). i.e; the classical Poisson bracket  algebra is retrieved by taking single contractions in the Operator Product Expansion of the quantum algebra. 
One can evaluate all expressions in the $r=\infty$ limit ( and set $d(t) =1$ for convenience.
The expectation value in the classical limit $<\rho|{\hat W}_2 ({\hat \rho}) |\rho>(\varphi (t))$
gives in the $r=\infty$ limit, after the dimensional reduction and after using  the asymptotic limits :

$${\partial \rho \over \partial r} =\varphi.~~~{\partial^2 \rho \over \partial r^2} = {\partial^2 e^\rho \over \partial
t^2}\rightarrow 0.~r\rightarrow \infty\eqno (4.14)$$
the value :

$$lim_{r\rightarrow \infty }<\rho|{\hat W}_2|\rho> = \int^{t_o}dt_1 \varphi (t_1) \int^{t_o}dt_1 \varphi (t_1).
\eqno (4.15)$$ 

after the normalization condition is chosen :

$$<\rho'|\rho>=\delta (\rho'-\rho).~<\rho|\rho> =1               \eqno (4.16) $$
We notice that eq-(4.15) is  the same as the integrand (3.9); so integrating 
(4.15) with respect to $t_o$ yields the energy
as expected.

It is useful to recall the results from ordinary $2D$ conformal field theory : given the  holomorphic current generator of
two-dimensional conformal transformations, $T (z)=W_2(z)$, the mode expansion is :

$$W_2 (z) =\sum_m~W^m_2 z^{-m-2}\Rightarrow W^m_2 =\oint~{dz\over 2\pi i} z^{m+2-1}W_2 (z). \eqno (4.17)$$
the closed integration contour encloses the origin. When the closed contour surrounds
$z=\infty$. This requires performing the conformal map $z\rightarrow (1/z)$ and replacing :

$$z\rightarrow (1/z).~dz\rightarrow (-dz/z^2).~W_2(z) \rightarrow (-1/z^2)^2 W_2 (1/z) =W_2(z)+{c\over 12}S[z',z]  \eqno (4.18)$$
in the integrand. $S[z',z]$ is the Schwarzian derivative of $z'=1/z$ w.r.t the $z$ variable.

There is also a $1-1$ correspondence between local fields and states in the Hilbert space :

$$|\phi> \leftrightarrow lim_{z,{\bar z}\rightarrow 0} {\hat \phi} (z,{\bar z}) |0,0>. \eqno (4.19)$$

This is usually referred as the $|in>$ state. A conformal transformation $z\rightarrow 1/z; {\bar z} \rightarrow 1/{\bar z}$:
defines the $<out|$ state at $z=\infty$

$$<out| =lim_{z,{\bar z}\rightarrow 0} <0,0|{\hat \phi} (1/z,1/{\bar z})(-1/z^2)^h (-1/{\bar z}^2)^{\bar h}  . \eqno (4.20)$$

where $h,{\bar h}$ are the conformal weights of the field $\phi (z,{\bar z})$.

The analog of eqs-(4.20) is to consider the state parametrized by $\varphi (t),d(t)$ :
$$|\rho>_{\varphi (t),d(t)} =lim_{ r\rightarrow \infty}~| \rho (r,t)>\equiv |\rho (out)>.$$      
 $$  |\rho>_{-\varphi (t),d(t)} =lim_{ r\rightarrow -\infty}~| \rho (r,t)>\equiv |\rho (in)>.\eqno (4.21)$$
since the continuous Toda equation is symmetric under $r\rightarrow -r\Rightarrow \rho (-r,t)$ is also a solution and it's
obtained from (3.1) by setting $\varphi \rightarrow -\varphi$ to ensure convergence at $r\rightarrow -\infty$.
As we pointed out earlier, the state $|\rho>$ is parametrized in terms of 
$\varphi (t),d(t)$ and for this reason one should always write it as
$|\rho>_{\varphi (t),d(t)}$ . What is required now is to establish the correspondence (a functor) between the representation space realized
in terms of the continuous Toda field and that representation ( the Verma module) built from the highest weight $|\lambda>$ 

$$<\lambda| {\tilde W} (D) |\lambda>={\tilde \Delta}^\lambda_1 \equiv \Delta^\lambda_2  \leftrightarrow 
<\rho|{\hat W}_2[{\hat \rho}(r,t)|\rho>.\eqno (4.22)$$

If one were to extract the zero mode piece of the ${\tilde W}_2$ operator via a contour integral around the origin , then integrate w.r.t. the $t$ variable and , finally, to  evaluate the expectation value,  one would arrive at a trivial  result.  There is a subtlety due the dimensional reduction from $2+1\rightarrow 1+1$ dimensions. If one $naively$ wrote down the expression   
  :

$$ {\cal P}[{ \Delta}^\lambda_1 ,{\bar  \Delta}^{\bar \lambda}_1]  \leftrightarrow 
<W^{0,0}_2>=\int^{2\pi}_0 dt_o~
_{\varphi}<\rho|[\oint{ dz\over 2\pi i }\oint{d{\bar z}\over
2\pi i} {\hat W}_2 (\rho(z,{\bar z},t))] |\rho>_{\varphi}. \eqno (4.23)$$ 
where one needs to add also the weights associated with the antichiral ${\bar W}_{\infty}$ algebra and the  real-valued generator  $W_2$ depends  on $r=z+{\bar z}$ and $t$ only. The l.h.s of (4.23) involves a dimensional reduction process in the weight space of the $W_\infty \oplus {\bar W}_\infty$ algebra which shall be discussed shortly. 

One now would be very hard pressed to avoid a zero answer after the contour integrations are completed in the r.h.s of (4.23). Expanding in powers of $(z+{\bar z})^n$ for negative and positive $n$ will give a trivial zero answer for the zero-mode of the $W_2$ operator. If one instead  expanded in positive powers of $(1/z+1/{\bar z})^n$; i.e. lets imagine 
expanding the function $e^{1/z}e^{1/{\bar z}}$ containing an essential singularity at the origin , in suitable powers of $z,{\bar z}$, the terms containing 
$z^{-1}{\bar z}^{-1}$ are the ones giving the residue. But $1/z +1/{\bar z} =(z+{\bar z})/z{\bar z}=r/z{\bar z}$ 
and this would contradict the assumption that all quantitites depend solely on the combination of $r=z+{\bar z}$ and $t$. Setting $t=z{\bar z}$ is incorrect because that will constrain the original $2+1$ Toda theory : $t$ is a parameter that appears in continuum graded Lie Algebras and has $nothing$ to do with the 
$z,{\bar z}$ coordinates. It plays the role of an extra ( compact ) coordinate, say an angle variable [4] but should not be confused with the $z,{\bar z}$ coordinates.

The correct procedure is to evaluate the generalization of what is meant by  eq-(4.17). The contour integral  means evaluating quantities for fixed times, which in the language of the $z,{\bar z}$ coordinates, implies choosing circles of $fixed$ radius around the origin and  integrating  w.r.t the angular variable. Therefore, the conserved Noether charges ( the Virsoro generators in the string case ) are just the  integrals of the conserved currents at fixed contour-radius ( fixed-times ). The  equal-time spatial ``hypersurfaces'' are then  the circles of fixed radius. The expression to evaluate is no longer (4.23) but 
the expectation value, say w.r.t the `in'' state, of the zero modes of the quantity  :

$${\cal L}_f= 
 \int^{2\pi}_0 dt'~f(t')~lim_{r'\rightarrow -\infty}~{\hat W}_2[\rho (r',t')]. \eqno (4.24)$$
where $f(t)=\sum_n a_n cos (nt)+b_n sin (nt)$. Hence, by zero modes one means 
those w.r.t the angle variable $t$ and $not$ w.r.t the $z{\bar z}$ variables. The zero mode generator is now given by its  $n=0$ value whose expectation value is  :
$${\cal P}[{ \Delta}^\lambda_1, {\bar  \Delta}^{\bar \lambda}_1] = <{\hat L}_{f_0}>= 
_{-\varphi(t)}<\rho |[ \int^{2\pi}_0 dt'a_0~lim_{r'\rightarrow -\infty}~{\hat W}_2[{ \hat \rho} (r',t')]]|\rho>_{-\varphi (t)}. \eqno (4.25)$$
In the l.h.s of (4.25) one may take  a suitable real-valued linear combination of the weights. There are two independent combinations involving  :

$${a_1\over 2}[{ \Delta}^\lambda_1 +{\bar  \Delta}^{\bar \lambda}_1]
+{a_2\over 2i}  [{ \Delta}^\lambda_1 -{\bar  \Delta}^{\bar \lambda}_1]. \eqno (4.26a)$$
where $a_1,a_2$ are two real coefficients and one imposes the reality constraint ${\bar \lambda }=\lambda^*,
{\bar \Delta}^{{\bar \lambda}}_1=({\Delta}^{\lambda}_1)^*$. What criteria will select the values of $a_1,a_2$ ? The most natural  dimensional reduction condition can be based on the fact that (4.12) was obtained simply by replacing 
$\partial/\partial z \rightarrow \partial/\partial r; \Theta \rightarrow \rho$ and a similar procedure with the antichiral generators. Therefore, we shall    
impose  the following reduction conditions for  the chiral/antichiral weights : 

$$ ({ \Delta}^\lambda_1)^* ={\bar  \Delta}^{\bar \lambda}_1.~{\bar \lambda}=
(\lambda)^*.~\Delta^\lambda_1 (\lambda) \rightarrow \Delta_1 ({\cal R}e~\lambda).~{\bar \Delta}^{\bar \lambda}_1 \rightarrow {\bar \Delta}_1 ({\cal R}e~\lambda^*)  \eqno (4.26b)$$
Per example, in eq-(4.10) the infinite number of weights is listed for a particular case, so  one would replace the complex variable $\lambda$ by its real component
, ${\cal R}e~\lambda$. Exactly the same is done  with the antichiral weights. If the coefficients of the weight polynomials had been complex-valued then one 
selects out their real components  as well. Eq- (4.26b) will be from now on our choice of the dimensional-reduction conditions in the weight spaces to yield the corresponding weights of the $U_\infty$ algebra.

The infinitesimal transformation law for the ${\hat \rho} (r,t)$ operator  under the 
${\hat L}_f [{\hat \rho} (r,t)]$ transformations  is :

$$\delta^\epsilon _{L_f}~{\hat \rho }(r,t) = 
 \int^{2\pi}_0 dt'~\epsilon f(t'){\hat W}_2[{\hat \rho} (r'\rightarrow \infty,t')].{\hat \rho}(r,t).         \eqno (4.27)$$
The Lie-algebra of derivations   of two infinitesimal $L_{f_1 (t)},L_{f_2 (t)}$ transformations is :

$$[\delta^{\epsilon_1}_{L_{f_1}}, \delta^{\epsilon_2}_{L_{f_2}}]= \delta^{\epsilon_3 (\epsilon_1,\epsilon_2)}_{[L_{f_1},L_{f_2}]}.~\epsilon_3 (t)\equiv \epsilon f_1(t){df_2\over dt}- \epsilon f_2(t){df_1\over dt} . \eqno (4.28)$$

This follows from the Virasoro-like transformations in (4.27). 
In (4.24) the $W_2$ generator has been used. A generalization to the other
$W_3,W_4,.......W_s...$ generators is straightforward and (4.27,4.28) are also valid due to the Lie-Poisson-Moyal nature of the original $W_\infty$
algebras/ area-preserving diffeomorphisms  :$[{\cal L}_f,{\cal L}_g]={\cal L}_{\{f,g\}}$ where the Moyal bracket must be taken between the $f,g$ functions w.r.t the internal coordinates $q,p$ of the two-dimensional surface, a sphere, plane, cylinder,.... per example. The area-preserving diffs of the plane is the classical $w_\infty$ algebra. For a cylinder is the $w_{1+{\infty}}$ algebra  and for the sphere is $su(\infty)$. Locally the algebras are isomorphic but $not$ globally. The $U_\infty$ 
symmetry algebra acting on the Toda molecule, stems from the $SU(\infty)$ Supersymmetric Gauge Quantum Mechanical Model associated with the light-cone gauge of the spherical supermembrane : a dimensionally-reduced super YM theory  to $one$ temporal dimension. The membrane's time coordinate ( $X^+$)  has a $correspondence $ with the   $r=z+{\bar z}$ variable. The extra coordinate arises from the $t$ parameter so the initial $3D$ continuos  ( super ) Toda theory 
is dim-reduced to  a $1+1$ ( super ) Toda $molecule$ : $\Theta (z,{\bar z},t)\rightarrow \rho (r,t)$ and, in this way,  an $effective$ two-dimensional theory emerges. Hence, the intrinsic $3D$ membrane theory can be related to  a ( non-critical) $2D~W_\infty$ string theory.   The algebra (4.28) is in effect a ``$Diff (S^1)$''-like algebra ( Virasoro-like) which can be extended, once we include $all$ the other $W_3,W_4,...W_s...$ generators,  to the $U_\infty$ transformations  of the whole $r,t$ two-dim space; in the same way that the $Diffs (S^1)$ algebra in closed-string theory can be  extended to the algebra of conformal reparametrizations of  the annular  ring in the punctured complex plane : the Virasoro algebra.   

We will discuss  what is meant by taking the expectation value w.r.t the ``in'' states. Firstly, if one is taking expectation values in the ``in'' states one has to evaluate the $lim~r'\rightarrow -\infty$ of the ${\hat W}_2$ generator. The latter algebra has to be  expressed in terms of the quantum solutions of the continuous Toda theory. Secondly,  
setting aside for the moment the  singularities appearing in  the OPE of Toda field  exponentials, prior to taking the $r'=-\infty$ limit,  one has after evaluating (4.25) :

$$_{-\varphi(t)}<\rho |{\hat I}(t=2\pi)-{\hat I}(t=0)|\rho>_{-\varphi (t)}=
_{-\varphi(t=2\pi)}<\rho |{\hat I}(t=2\pi)|\rho>_{-\varphi (t=2\pi)}-$$
$$ _{-\varphi(t=0)}<\rho |{\hat I}(t=0)|\rho>_{-\varphi (t=0)}=I(2\pi)-I(0). \eqno (4.29)$$
after using the normalization condition $<\rho|\rho>=1$. Thirdly, modulo
the subtleties in pulling the expectation value inside the $t'$-integration,  
one can verify that the same  answer is obtained if one performs the expectation value before the $t'$ integration . One evaluates the action of an operator of $\rho$ acting on the ``in'' state as follows :

$${\cal F}[{\hat \rho}(r',t')] |\rho>_{-\varphi (t)}=lim_{r'\rightarrow -\infty,t'\rightarrow t}~{\cal F}[\rho (r',t')]|\rho>_{-\varphi (t)}. \eqno (4.30)$$ 
By inspection one can verify that after performing the integration one arrives at the same  (4.29) . This is the anlog of $f[{\hat x}]|x'>=f(x')|x'>$. The integral (4.24) clearly diverges as a result of the OPE of the Toda exponentials prior to taking  the $r'=-\infty$ limit. This can be fixed by introducing a ``normal ordering'' procedure that removes the short distance singularities as a result of the coincidence limit of two or more operators.  What is important is the $t$ integration. Since the l.h.s of (4.25) has a dependence on the $\lambda,{\bar \lambda}$ parameters the $\varphi (t)$ must also encode such a dependence. The spectral decomposition of $\varphi (t)$ can be taken to be :

$$\varphi_{\lambda, {\bar \lambda}} (t) =\sum_n A_n (\lambda,{\bar \lambda})sin(nt) +B_n (\lambda,{\bar \lambda})cos(nt). \eqno (4.31)$$
plugging (4.31) into (4.25) yields one equation with an explicit 
$\lambda,{\bar \lambda}$ dependence on both sides of the equation.
The other infinite number of integrals like (4.25) involving the other weights associated with the zero modes of the algebra generators are :

$${\cal P}[{ \Delta}^\lambda_k , {\bar  \Delta}^{\bar \lambda}_k]   \leftrightarrow 
_{-\varphi(t)}<\rho |[\int^{2\pi}_0 dt'~a_0
{\hat W}_s[{\hat \rho} (r'=-\infty,t')]]|\rho>_{-\varphi (t)}. \eqno (4.32)$$
with $k=2,3,4......\infty$ and $s=k+1 =3,4,.....\infty$ and the coefficient $a_0$ will be determined in ${\bf 4.3}$.
The infinite number of equations defines the ``coefficients'' appearing in the spectral decomposition of $\varphi (t)$, $A_n (\lambda, {\bar \lambda}),
B_n (\lambda, {\bar \lambda})$. 

Let us suppose that the weights given by (4.10) were the ones chosen  and the criteria given in (4.26b) to set the dimensional-reduction condition was used for the l.h.s of (4.32) . The weights are then  polynomials in ${\cal R}e~\lambda $. Choosing the coefficients $A_n (\lambda, {\bar \lambda}),
B_n (\lambda, {\bar \lambda})$ to be polynomials of order $n$ in these variables will be sufficient to solve for (4.25,4.32). A real-valued polynomial of order $n$ has $n+1$ independent coefficients. Say one opts to choose : $A_n (\lambda, (\lambda)^* )=a^n_o +a^n_1\lambda +a_2\lambda^2......+a^n_n (\lambda)^n  +c.c$. Each real-valued polynomial of order $n$ has $n+1$ independent coefficients, thus, the total number of independent coefficients in the infinite number of polynomials   is of order 
$\sum_{n=1}^\infty 2(n+1)\sim n^2$, where one has  included  those coefficients stemming from the  $B_n$ terms as well. Now, the l.h.s of (47,54) generates $k$ equations, each having  the weight-polynomials  of degree  $k+1$ in $\lambda$  containing $(k+2)$ known coefficients;  so the total number of known coefficients is $\sum_1^\infty (k+2)\sim k^2/2$.   Since $(2\infty)^2=(\infty)^2$ the cardinality matches so that one does not end up with an undetermined system of equations. Therefore, by matching, level by level in $k$, the coefficients involved in the powers of $\lambda^0,\lambda^2,\lambda^3,......$ one can solve for the $A_n,B_n$ coefficients and determine $\varphi (t)$ in (4.31). 

Therefore, a knowledge of the weights associated with a given representation fixes the form of the function $\varphi (t)$ and furnishes the $|\rho (in)>_{-\varphi (t)}$ state uniquely in terms of the $\lambda, {\bar \lambda}$ parameters. The same applies to the $|\rho (out)>_{\varphi (t)}$ state. Thus, the 
quantum states of  the continuous Toda theory  can be  explicitly classified  in terms of 
the representations of the $W_{\infty}\oplus {\bar W}_{\infty}$ $after$ the dimensional reduction is taken place.

This is not the end of the calculation. We still have to go back to the original quantum version of eqs-(2.4) in order to have an exact quantization of the 
YM potentials that are in a $1-1$ correspondence with the membrane coordinates. We are assuming that the correspondence between the Toda theory and the YM theory also holds  off-shell. By  replacing Poisson brackets by commutators in (2.4), one can obtain the commutation relations of the YM operator-valued fields, ${\hat A}_y (\tau),{\hat A}_{{\bar y}}(\tau), {\hat A}_3 (\tau)$ in terms of the $\theta_l (\tau)$ fields. This was already discussed in ${\bf II}$  
 
The knowledge of the Hilbert space of states $|\rho_{\lambda,{\bar \lambda}} (in)>$ linked to the exact quantization of the continuous Toda field determines the Hilbert spaces associated with each one  of the infinite number of $\theta_l (\tau)$ fields that appear in the decomposition  : ${\cal H}_{\infty}={\cal H}_1\oplus 
{\cal H}_2 \oplus ......$. Hence,  
the spherical membrane in moving flat backgrounds ( the zero modes must be treated separately ) is an exact integrable quantum system insofar that it is equivalent to a $SU(\infty)$ YM theory dimensionally-reduced to one temporal dimension and the latter, in turn, can be re-expressed ( using the ansatz given by (2.4) ) in terms of an exact quantum integrable continuous Toda theory  ( with the  natural action of the  $U_{\infty}$ symmetry ). 

To conclude, the Hilbert space of states, $|\rho_{\lambda,{\bar \lambda}} (in)>$ can thus be  obtained ( in principle )  directly from eqs- (4.25,4.32) for any given highest weight irrepresentation of the $W_{\infty}$ algebra. This is the essence of this work. We must not forget that the symmetry algebra acting on the integrable system is the dimensionally- reduced $U_{\infty}$ algebra.  

\smallskip
\centerline {\bf 4.3 The $U_\infty$ Algebra in the $Z,{\bar Z}$ Representation }
\smallskip
So far we have worked with integrals involving the $r=z+{\bar z},t$ variables. One could 
define the $new$ complex 
variables exploiting the fact that $t$ is a $periodic$ variable parametrizing a circle  [4] :

$$U\equiv t+ir.~{\bar U}\equiv t-ir.~Z=e^{-iU}=e^{r}e^{-it}.~{\bar Z}=e^re^{it} \eqno (4.33)$$
to obtain  a new complex domain $Z,{\bar Z}$ from the cylinder defined by the 
$U$ coordinates such that $r=-\infty$  corresponds to the 
origin $Z,{\bar Z}=0$ and  $r=+\infty$ corresponds to infinity. In this fashion a Hamiltonian analysis based on $r$ quantization ( the original time coordinate) corresponds to  a 
``radial'' quantization in the complex plane as  conventional CFT requires.  It is precisely when this correspondence is made that the two-dimensional world of non-critical $W_{\infty}$ strings matches the two effective dimensions which follow the dimensional reduction from the $2+1$ continuous  Toda theory to 
$1+1$ dimensions ; that of the $r,t$ variables. It is in this fashion how it makes sense to claim that non-critical $W_{\infty}$ strings in $D=27$ are connected to critical bosonic membranes in $D=27$ ( after the light-cone gauge is chosen so a $SU(\infty)$ YM theory dimensionally reduced to one temporal dimension  emerges ). Both theories are effectively two-dimensional. $W_{\infty}$ string theory amounts to an infinite collection of Virasoro-like strings with an infinite number of unusual central charges and intercepts ( linked to Regge-like trajectories ) [25]  . That infinite collection of strings effectively behaves like a membrane ( excluding the membrane's zero modes ). One should not worry at the moment about the fact that the area-preserving-diffs of the cylinder is $w_{1+{\infty}}$ algebra instead of the $w_\infty$ algebra associated with the plane. The presence of the $w_\infty$ structures orginate from the $3D$ continuous Toda theory. The physically relevant area-p!
reserving-diffs algebra which one is always referring to is the $su(\infty)$ algebra associated with the sphere.  

It is important to point out that the generator $W_2 [\rho (r,t)]$ (4.12) obtained as a dimensional reduction of (4.11) ( and its antichiral counterpart ) is a $mixed$ tensor w.r.t the $Z,{\bar Z}$ coordinates. This is easily derived by starting with  the integral of $W_2 [\rho (r,t)]$ w.r.t the ``angle'' variable $t$ to yield the energy. The  effective two dimensional space with coordinates, $r,t$, is a cylinder. $r$ plays the role of time and $t$ of space ( an angle ). Hence the conserved quantity w.r.t. the timelike Killing vector, $v^b = v^r(\partial/\partial r)$, is   
the energy given by the integral over a spacelike surface $\Sigma^a$ ( whose
oriented normal is timelike ) of the quantity : $T_{ab}v^b$ :

$${\cal H}=\int dt~W_2[\rho (r,t)]\equiv 
\int d\Sigma^aT_{ab}v^b =\int dt~T_{rr}  =   Energy. \eqno (4.34)$$ 
where $d\Sigma^r=dt;v^r=1$.
Because $T_{rr}(r,t)\equiv W_2 [\rho (r,t)]$ is one component of the stress energy tensor it must transform under the coordinate transformations : $r\rightarrow {1\over 2}ln(Z{\bar Z}).~t\rightarrow {1\over 2}ln(Z/{\bar Z})^i$ as follows :

$$T_{rr} (r,t)=W_2 [\rho (r,t)]={\partial Z\over \partial r}
{\partial Z\over \partial r}W_{ZZ} (Z,{\bar Z})+
{\partial {\bar Z}\over \partial r}{\partial {\bar Z}\over \partial r}
W_{{\bar Z}{\bar Z}} (Z,{\bar Z})+$$
$${\partial Z\over \partial r}
{\partial {\bar Z}\over \partial r}W_{Z{\bar Z}} (Z,{\bar Z})+
{\partial {\bar Z}\over \partial r}
{\partial Z\over \partial r}W_{{\bar Z}Z} (Z,{\bar Z}). \eqno (4.35)$$
An important remark is in order. Imagine knowing the remaining components 
$T_{rt},T_{tt},T_{tr}$ of the stress energy tensor in addition to $T_{rr}\equiv
 W_2[\rho (r,t)]$. In terms of the new complex coordinates a new tensor is constructed with components :$W_{ZZ};W_{{\bar Z}{\bar Z}}....$. Does this stress energy tensor generate new holomorphic/antiholomorphic transformations ? $Z\rightarrow Z'(Z);{\bar Z}\rightarrow {\bar Z}'({\bar Z})$ The answer is $no$. By inspection we can see, per example, that under dilatations  $z'(z)=\lambda z;{\bar z}'({\bar z})=\lambda {\bar z}$
 one does not generate $ Z'(Z)=\Lambda Z;{\bar Z}'({\bar Z})=\Lambda {\bar Z}$. And viceversa, in general $F(Z),{\tilde F}({\bar Z})$ does not always imply 
 $z'=f(z); {\bar z}'={\tilde f}({\bar z})$. Thus, the components of the new stress energy tensor cannot longer be split in terms of two $independent$  holomorphic/antiholomorphic pieces : $T(Z)=W^+_{(2,0)}(Z),{\tilde T}({\bar Z})=
W^-_{(0,2)}({\bar Z})$ as it occurs in ordinary CFT. The dimensional reduction $intertwines$ the holomorphic/antiholomorphic variables and the dimensional reduced 
$W_{\infty}\oplus {\bar W}_{\infty}$ algebra, is the symmetry  algebra $U_{\infty}$ acting on the Toda molecule,  not to be confused with the $V$ algebras constructed by Bilal. The $U_\infty$ algebra does $not$ generate conformal transformations in the $Z,{\bar Z}$ variables.

We can write down eqs-(3.9,4.15) again for the classical generator $W_2$   and by inspection  see that :

$$W_2[\rho (r,t)]\sim (\int^tdt'~\varphi (t'))^2=F(t)=F[(ln~(Z/{\bar Z})^i]\not= f(Z)+{\bar f}({\bar Z}). \eqno (4.36)$$
A splitting cannot occur unless a very special choice of $\varphi (t)$ is chosen . There is no reason why $F(U+{\bar U})=f(U)+{\bar f}({\bar U})$ unless one has a linear function. Another way of proving that there cannot be a split is by
computing :

$$<\lambda,{ \bar \lambda}| \oint {dZ\over 2\pi i}T(Z)|\lambda,{ \bar \lambda}>=
<L_{-1}>=0.~
<\lambda,{ \bar \lambda}| \oint {d{\bar Z}\over 2\pi i}{\tilde T}({\bar Z})|\lambda,{ \bar \lambda}>=
<{\tilde L}_{-1}>=0. \eqno (4.37)$$
where one has used the fact that the spurious sates 
$S=L_{-1}|\phi>,{\tilde S}={\tilde L}_{-1}|\phi>$ are orthogonal to the 
physical states, $|\phi>$. It is now fairly clear that the $<{\hat W}_2[\rho (r,t)]>$ cannot be expressed in terms  of the two expectation values given by (4.37). One cannot claim that the r.h.s of eq- (4.35) 
contains solely the $T(Z),{\tilde T}({\bar Z})$ pieces as ordinary CFT does.
The same argument applies to the terms in the r.h.s of (4.35) :$Z^2W_{ZZ},
 {\bar Z}^2 W_{{\bar Z}{\bar Z}}$. If one (wrongly) assumes that the mixed components are zero ( like in ordinary CFT) and that $W_{ZZ}$ is purely holomorphic and $W_{{\bar Z}{\bar Z}}$ is antiholomorphic,  after taking expectation values of the $L_1,{\tilde L}_1$ generators like it was performed in (4.37)  one gets zero. This wouldn't agree with $<W_2[\rho (r,t)]>$.
Shortly an explicit expression for the components of the r.h.s of (4.35) will be given.

Let us define a meromorphic real-valued field operator of two complex variables, $Z,{\bar Z},~{\hat W}_2 (Z,{\bar Z})$, to be determined in terms of $W_2[\rho (r,t)]$ admitting the following expansion in powers of $Z,{\bar Z}$ :

$${\hat W}_2 (Z,{\bar Z})=\sum_{N,{\bar N}}~{\hat W}^{N,{\bar N}}_2
Z^{-(n+im)-1}  {\bar Z}^{-(n-im)-1}. \eqno (4.38) $$
where $N\equiv n+im.~{\bar N}\equiv n-im$. The  mode expansion in the $Z,{\bar Z}$ variables (4.38) $differs$ from the expansion w.r.t the $z,{\bar z}$ variables as a result of the fact that the generator  $W_2[\rho (r,t)]=T_{rr} (r,t)$ is a particular component of a  $mixed$ stress energy tensor tensor  that intertwines $z,{\bar z}$. This is to be expected. 
Therefore, the correct expression to extract the residues is :

$${\cal P}[{ \Delta}^\lambda_1,{\bar  \Delta}^{\bar \lambda}_1] \leftrightarrow <\rho (in)|{\hat W}^{N=0,{\bar N}=0}_2 |\rho (in)>= 
<\rho (in)|[\oint{ dZ\over 2\pi i }\oint{d{\bar Z}\over
2\pi i}  {\hat W}_2 (Z,{\bar Z}) |\rho (in)>. \eqno (4.39)$$

If (4.39) is equal to (4.25) then certain conditions must be met . If we integrate (4.39) around circles of fixed-radius $R=e^r$ ( and then take the $R\rightarrow 0$ limit) one has after using : $dZ=-iZdt;d{\bar Z}=i{\bar Z}dt$ and taking into account an extra minus sign due to the counterclockwise/clockwise $Z,{\bar Z}$ contour integrations the following condition :

$$\int^{2\pi}_0{dt\over 2\pi}~\int^{2\pi}_0{dt\over 2\pi}~lim_{R\rightarrow 0}R^2{\hat W}_2 (Re^{-it},Re^{it})=\int^{2\pi}_0dt~a_0W_2[\rho (r=-\infty,t)]. \eqno (4.40)$$

hence one learns from (4.40) and (4.35) that :
$$a_0={1\over 2\pi}.~W_2[\rho (r,t)]=Z{\bar Z}{\hat W}_2 (Z,{\bar Z})=Z{\bar Z}(W_{Z{\bar Z}}+
W_{{\bar Z}Z}).~Z^2W_{ZZ}+{\bar Z}^2 W_{{\bar Z}{\bar Z}}=0. \eqno (4.41)$$

Later we shall study the conditions on the $f(t)=\sum a_n cos (nt) +b_n sin (nt)$ appearing in (4.24) 
in order to find the relations amongst the higher order modes than the zero ones and for all the $W_s$ generators in addition to the $W_2$.

Despite the fact that one has no longer purely  holomorphic/antiholomorphic transformations, the transformations generated by the $U_{\infty}$ generators  
${\hat W}_2 (Z,{\bar Z}) $ acting on ${\tilde \rho}(Z,{\bar Z})$ can, nevertheless,  be written as :

$$\delta^\epsilon_{W_s}~ {\tilde \rho}=\oint \oint {dZ\over 2\pi i} 
{d{\bar Z}\over 2\pi i }\epsilon (Z,{\bar Z}){\hat W}_s (Z,{\bar Z}){\tilde \rho }(Z',{\bar Z}'). \eqno (4.42)$$
after using the standard techniques of OPE and contour deformations of CFT. The contours sorround the $Z',{\bar Z}'$ variables and $not$ the origin after the contour deformation is performed. The $U_{\infty}$ algebra obtained from the dimensional-reduction process of the $W_{\infty}$ extended conformal field theory  inherits 
a similar transformation structure.

Going back to (4.39), the $|\rho (in)>_{-\varphi (t)}$ state at $r=-\infty$ is now the state corresponding to the operator insertion at the origin of the punctured  $Z,{\bar Z}$ plane.
To define the ``in'' state requires evaluating the limit :

$$|\lambda,{\bar \lambda}>\leftrightarrow lim_{Z,{\bar Z}\rightarrow 0}~
r(Z,{\bar Z}) {\hat \varphi}_{\lambda,{\bar \lambda}} [t(Z,{\bar Z})]~|0,0>. 
\eqno (4.43)$$.

The operator  quantity $-{\hat \varphi} (t)$ appearing in the quantum Toda solutions bears an explicit $\lambda,{\bar \lambda}$ depenence ( to be determined below ) and defines  the $|in>$ state at  $r=-\infty$ in  eq-(4.21). Now it makes sense to evaluate the contour integrals using the residue theorem  without obtaining a trivial zero answer.

If one opts to perform the contour integration $before$ the expectation value is taken in (4.39) is more convenient to take a contour surrounding the 
$origin$ which will absorb the singularities of the OPE in the coincidence limits $Z',{\bar Z}'\rightarrow Z,{\bar Z}\rightarrow 0$.  In general the contour integration does $not$ necessarily  commute with the evaluation process of taking the expectation value :
$$<W^{0,0}_2>\sim  \oint{ dZ\over 2\pi i }\oint{d{\bar Z}\over
2\pi i} <\rho_o|[ {\hat W}_2 ((Z,{\bar Z})] |\rho_o>. \eqno 
(4.44)$$
because one cannot naively  pull out the contour integrals outside the expectation values without introducing contact terms due to the time-ordering 
procedures ( in  the  expectation values ) leading to delta function singularities. Setting these subtleties aside, integrals (4.39,4.44) are  roughly the same.

If there were no singularities in the OPE defining the quantum ${\hat  W}_2(Z,{\bar Z})$ generator then (4.39) would be zero; per example, in the quantum case,  the expectation value of  : $<\rho_o|{\hat W}_2 |\rho_o>$ has singularities as result of  coincidence limits arising in the OPE of Toda exponentials  and for this reason integrals like (4.39) are not trivial. The singularities occur as $r'\rightarrow r\rightarrow -\infty$ which impies $Z'\rightarrow Z\rightarrow 0$ ( similarly with the 
${\bar Z}$ coordinates ) in the punctured complex plane. 
In the proceeding paragraphs we shall discuss the need to regularize certain expectation values   in order to evaluate  the conserved quantum integrals (charges )  of motion whose expectation values do not depend on the ``time'' variable; i.e. the quantum equations of motion for the operators obey : $d{\hat I}_n/dr =0;n=2,3....$ and these charges are involutive  $[{\hat I}_n, {\hat I}_m ]=0$.

How does one know that the right singularity structure appears in the evaluation of (4.39)? . To begin with the continous Toda field is $not$ a primary field  ( like the stress energy tensor in ordinary CFT)  and singularities of the required type do appear. In particular, the OPE of the exponentials of the ordinary Liouville field has been computed in great detail by Gervais and Schnittger [31] and the underlying quantum group structure was discovered associated with the algebra of chiral vertex operators. Savaliev has shown that a  realization of the $W_{\infty}\oplus {\bar W}_{\infty}$ algebra can be given $precisely$ in terms of the $3D$ continuous Toda field, and, $as~ such$,  one must have, accordingly, the correct singularity structure of the OPE in order to evaluate 
eq-(4.39). This is what a faithful realization of the $W_{\infty}$ entails.  Analogous to the results of [31],  
the OPE of the continuous-Toda fields exponentials should have singularities in the form of negative  
powers in $(Z-Z')^{-N}({\bar Z}-{\bar Z}')^{-{\bar N}}$ and the contour integrals
around the origin in  (4.39) can be computed yielding a non-zero result.

The same arguments can be applied to the $U_{\infty}$ algebra. The dimensional reduction shouldn't spoil ( in principle ) the presence of the singularities which appeared in the original $3D$ theory. One is free to work always in the $r,t$ representation where the equal time surfaces are the cylinders and $t$ is the angular variable. 
If one is worried about these technicalities one may quantize the classical $U_{\infty}$ algebra in the $r,t$ space by recurring to the explict quantum solutions (3.1.3.6,3.7) and the use of OPE based on the $r,t$ variables. If reparametrization invariance is not spoiled by the quantization process the results should be independent on which set of variables one decided to use. If the original $3D$ Toda theory is anomaly free the dimensional reduction process should be as well. The converse is not true. There is no reason why the supermembrane in $D=11$ is anomaly free just  because the $D=10$ superstring is.    

In (4.39) the states $|\rho (in)>$ are the sought-after highest weight states (ground states). The physical states belong to a particular class of the former. In order to obtain the space of physical states a complete list of all $unitary$ highest weight irrepresentations is required and from these a no-ghost theorem can be formulated which will select the restricted values of the infinite number of conformal weights
, $\Delta_k$, as well as selecting the value of the critical central charges ( or spacetime dimension as it occurs for the string ). This is tantamount of writing down the complete BRST cohomology 
of the physical vertex operators linked to the ( dimensionally-reduced ) $W_{\infty}$ CFT. This is a extremely arduous task because we don't have a $W_{\infty}$ CFT. Ordinary string theory is based on a $W_2$ CFT, rational or irrational CFT, and we know how difficult matters can be.  From each of these sought-after highest weight states
one builds a tower (the Verma module) by succesive applications of the ladder like operators (4.5). In ordinary string theory the tower of states are the so called spurious states. These are $physical$
when $D=26,a=1$ because the norm is zero. Eq-(4.39) is the analog of the string mass shell condition :$(L_o -a)|\phi>=0$ where $a=1$ results from the
zeta function regularization of the string zero-point-energy states. This is the reason why the expectation values of the zero mode operators $W_2^{N,{\bar N}}$ would require regularization as well. 

The computation of (4.39) is messy. Firstly, in order to compute the quantum analog of  eqs-(4.12) by replacing the classical continuous Toda solutions by the quantum operators,   would require to develop the OPE of Toda exponentials as well  as the  extension of the quantum enveloping algebra $U_q [sl(\infty)]$ associated with the vertex operator algebra; this hasn't been performed according to our knowledge.
This is the $major$ obstacle in obtaining the  exact quantization of the  membrane with the topology of a sphere. Once the OPE of the continuos Toda field exponentials is constructed
the tools to furnish  highest weight irreducible representations in terms of solutions of the quantum continuos Toda theory should be enough to quantize 
$exactly$ the spherical membrane; i.e. to build the fully fledged non-perturbative spectrum in apropriate ``$U_{\infty}$'' invariant  backgrounds.  The no-ghost theorem based on the unitary representations can then be studied to select critical spacetime dimensions. This is tentative ,of course. The main point is that the spherical membrane is an exact quantum integrable model and hence solvable. In the next section  simplified ways to solve this problem are  presented that allows to bypass the  explicit computation of the OPE. Thus, exact results can be  obtained. 

If the quantization of (4.12) displays singularities in the operator products so will the quantization of (4.15) and hence the energy is ill-defined. Therefore, a regularization is required so that the expectation value of (4.12) in the quantum case is finite. In ordinary CFT this is achieved  after a conformal normal ordering procedure ( or other suitable ordering ) is introduced so that correlations functions are finite and obey the Ward identities. The conformal normal ordering removes in most cases the infinities. It is customary to define the  normal ordered  product of two field-operators, $({\cal A}{\cal B})(w)$  as a contour integral surrounding $w$ of $(z-w)^{-1}{\cal A}(z){\cal B}(w)$. It is essentially a point-splitting regularization method.  The normal ordering does not obey the commutative nor associative property.         

With this lesson in mind let us introduce now the  family
of functions $\varphi_{\lambda,{\bar \lambda}} (t),d_{\lambda, {\bar \lambda}} (t)$ which appear in the general solution to the quantum continous Toda equation given by eqs-(3.1,3.6,3.7) ( to be determined below ) parametrized by $\lambda,{\bar \lambda}$; i.e to each 
highest weight irrepresentation parametrized in the form :$|\lambda,{\bar \lambda}>$
one associates the family of functions of $t$ parametrized by $\lambda,{\bar \lambda}$. We shall omit the suffixes in $\varphi (t)$ for the time being.
Due to the exact quantum integrability of the continuous Toda theory, the 
$<\rho (in)|{\hat W}_2|\rho (in)>$, after regularization, must have the same functional dependence  on  $\varphi (t)$ as the one given by the classical energy eq-(3.9). 
Inotherwords,  upon quantization a physical regularization prescription  must be one  such that the expectation value in the $|in>$  states   : 
 
$$  <\rho (in)|{\hat W}_2|\rho (in)>_{reg} = \int^{t_0}dt_1~ \varphi (t_1)
\int^{t_0}dt_1~\varphi (t_1). \eqno (4.45)$$
should agree with expression (3.9). The regularization method involves the point-splitting method discussed above. 
$\varphi (t)$ must now be interpreted as the expectation value ( to be determined shortly ) of the ${\hat \varphi} (t)$ operator which appears in the exact quantum solutions (given  by eqs-(3.1,3.6,3.7). 
These asymptotic expressions are given in terms of eq-(3.4) where now  $\varphi (t),ln~d(t)$ are operators obeying the commutation relations (3.8b). To simplify matters we choose $d(t)$ to be the unit operator so (3.8b)  constrains the $c$-number function $w(t)$ to satisfy 

$${1\over w(t)}{d^2 \over dt^2}  ({1\over w(t)})=0. \eqno (4.46)$$
so that eq-(3.7) has only an  $\hbar$ term in the delta function term.

The state $|\rho>$ in general 
 could be any state in the Hilbert space of states  associated with the operator ${\hat \rho}(r,t)$ for any value of $r$; i.e. it is the state associated with the $interacting$ highly nonlinear Toda field. At the moment we don't know if there is a $1-1$ correspondence between local fields inserted at particular locations in the punctured complex plane ( or other Riemann surface ) and states in this dimensionally-reduced $W_{\infty}$ CFT, as it happens with ordinary CFT. Choosing the asymptotic  states  simplify matters considerably  because the fields become free.  The asymptotic limit  was  uniquely determined in eq-(3.4) by the operator 
${\hat \rho}_o=(\partial^2 {\hat x}_o/\partial t^2)\rightarrow r{\hat \varphi} (t)$. 

A question comes to mind :``Where does  the $ \hbar$-dependence of the energy 
 ( given by the integral of eq-(4.15) w.r.t the  $t$ variable ) come from ?'' In eqs-(3.6,3.7)
the quantization was encoded, for the most part, in the function 
$w(t)$ which yields the $O( \hbar)$ corrections to the $\sum\varphi(t)^{-1}$ terms in eq-(3.1). In this  Heisenberg representation the quantum solutions for $\rho (r,t)$ are seen as operator valued quantities with an $explicit$ $\hbar$ dependence in (3.6,3.7). It is for this reason that $\varphi (t)$ must be  taken to be an operator ( as well as $d(t)$). After expectation values are taken in the asymptotic limits  the  explicit $\hbar$ factors appear in the eigenvalues of the operator ${\hat \varphi (t)}$ that  carry the $ \hbar$ dependence.  Whence the presence of the $ \hbar$. ( $d(t)$ was chosen  earlier to be the unit operator ). The eigenvalue equation needed to compute the expectation values in the 
$|\rho (in)>$ states  reads :

$$|\rho_{\lambda, {\bar \lambda}}> =lim _{Z,{\bar Z}\rightarrow 0} 
~r(Z,{\bar Z}){\hat \varphi}_{\lambda, {\bar \lambda}} [t(Z,{\bar Z})]~
|0,0>. \eqno (4.47)$$                  
The $ \hbar $ dependence is encoded in the eigenvalues/ weights. Per example, the angular momentum states  are $|J,J_z>$ such that 
${\hat J}_z |J,J_z>=m_z \hbar |J,J_z>$

Integrals like (4.39) are not the only ones required to determine 
$\varphi_{\lambda, {\bar \lambda}} (t)$
To find such an explicit $\lambda,{\bar \lambda}$ dependence of $\varphi (t)$ one needs to recur to $all$ the  weights of the representation, 
$  \Delta^\lambda_k$  ( and the
anti-chiral ones ). These  are also related to  the zero modes of Saveliev's realization of the chiral and antichiral $W_{\infty}$
algebras in terms of the dressed continous Toda field [4], where the infinite number of generators 
have similar  form  to eq-(4.11)  ( the number of  $t$ integrations depends on the value of the conformal spin )  : 

$${\cal W}^+_{(h,0)} [\partial \rho/\partial z;....\partial^h \rho /\partial z^h].
~{\cal W}^-_{(0,{\bar h})}
[\partial_z \rho \rightarrow \partial_{{\bar z}}\rho].~\partial {\cal W}^+/\partial {\bar z} =0.~\partial {\cal W}^-/\partial  z =0\eqno
(4.48) $$.  
After the dimensional reduction is taken, the expectation values of the zero modes,  expressed  in terms of the new variables, $Z,{\bar Z}$, of the $U_\infty$ generators    are :
$$<|{\hat W}^{N=0,{\bar N}=0}_s |>=
<\rho (in)|[\oint~{dZ\over 2\pi i}\oint {d{\bar Z}\over 2\pi i}   Z^{{s\over 2}-1}{\bar Z}^{{s\over 2}-1} {\hat W}_s (Z,{\bar Z})] |\rho (in)> 
\leftrightarrow {\cal P}[ \Delta^\lambda_k ,{\bar \Delta}^{\bar \lambda}_k]  . \eqno (4.49)$$
where $k\ge 1,~k=2,3,4......\infty$. The real-valued meromorphic field operator, ${\hat W}_s (Z,{\bar Z})$,  admits an expansion similar to (4.38) in powers of $Z^{-N-s/2}{\bar Z}^{-{\bar N}-s/2}$. 
We notice again a difference  between the exponents of $Z,{\bar Z}$ versus the usual ones in the $z,{\bar z}$ variables : $z^{h-1}{\cal W}^+_h;
{\bar z}^{{\bar h}-1}{\cal W}^-_{{\bar h}}$. 
The fact of having in some instances half-integer exponents in (4.49) is not that harmful; these are very natural in the fermionic string.

If eqs-(4.32) are to be equal to (4.49) similar conditions like (4.40) must be met. After performing the contour integrals around the origin by means of cirlcles of fixed radius  as it was done in (4.40) , 
using  $Z=Re^{-it},{\bar Z}=Re^{it},R=e^r$, yields :

$$ \int^{2\pi}_0{dt\over 2\pi }~lim_{R^2\rightarrow 0}(R^2)^{{s\over 2}}
 {\hat W}_s (Re^{-it},Re^{it}) =\int^{2\pi}_0dt~a_0
W_s[\rho (-\infty,t)]. \eqno (4.50)$$

this occurs for fixed radius, $R=e^r$, hence one learns that in general one must have  :

$$a_0={1\over 2\pi}.~(Z{\bar Z})^{s/2}{\hat W}_s (Z,{\bar Z})=W_s[\rho (r,t)]. \eqno (4.51)$$

The conditions on the other higher modes can be met also if the $n^{th}$ component, $f_n(t)$, of the function $f(t)$ obeys the following equation :

$$\int^{2\pi}_0dt~f_n(t)W_s[\rho (-\infty,t)] =\int^{2\pi}_0 {dt\over 2\pi}
lim_{R\rightarrow 0}~(R^{2})^{s/2}(R^2)^n (Z/{\bar Z})^{im}{\hat W}_s (Re^{-it},Re^{it}). \eqno (4.52)$$
where one $t$ integration absorbs a $2\pi$ factor.

An equality can be obtained iff $m={n\over 2} \Rightarrow N\equiv n(1+i/2);
{\bar N}\equiv n(1-i/2)$ and the $f(t)$ admits the expansion in $cosh~(nt); sinh ~(nt)$ instead of cosines/sines :
$f(t)=\sum_n a_ncosh~(nt)+b_nsinh~(nt)$. Plugging $f_n(t)$ in (4.52) yields after matching term by term in $n$ :

$$lim_{R\rightarrow 0}(R^2)^{s/2}{\hat W}_s(Re^{-it},Re^{it})=W_s[\rho (-\infty,t)]. $$ 
$$ a_ncosh~(nt)+b_nsinh~(nt)=lim_{R\rightarrow 0}~{R^{2n}e^{nt}\over 2\pi}. \eqno (4.53)$$
If the last relation holds at a $fixed$ value of $R=e^r$ and its generalization to other values of $Z,{\bar Z}$ this would require incorporating an $r$ dependence to the original $f(t)$ function so that the  $a_n (r),b_n(r)$ coefficients must  have an explicit $r$ dependence in order for (4.53) to be consistent.
The condition ( valid for other values of $Z,{\bar Z}$ ) is 

$$(Z{\bar Z})^{s/2}{\hat W}_s (Z,{\bar Z})=W_s[\rho (r,t)].~
a_n (r)={\tilde a}_n e^{2nr}.~b_n (r)={\tilde b}_n e^{2nr}. \eqno (4.54)$$
Eq-(4.53) yields :
$${\tilde a}_n \equiv lim_{R\rightarrow 0} {a_n (r)\over R^{2n}}. 
~{\tilde b}_n \equiv lim_{R\rightarrow 0} {b_n (r)\over R^{2n}}. \eqno (4.55)$$
so that the $R=e^r$ factors cancel out leaving  : 
$${\tilde a}_ncosh~nt+{\tilde b}_nsinh~nt={e^{nt}\over 2\pi}. \eqno (4.56)$$
in agreement with the previous conclusion that $a_0={\tilde a}_0 =(1/2\pi)$.

Concluding, we have that the real-valued meromorphic field 
${\hat W}_s (Z,{\bar Z})$ is related to $W_s[\rho (r,t)]={\tilde W}_s [{\tilde \rho}(Z,{\bar Z})]$ as follows :

$$(Z{\bar Z})^{s/2}{\hat W}_s (Z,{\bar Z})=W_s[\rho (r,t)]={\tilde W}_s [{\tilde \rho}(Z,{\bar Z})].\eqno (4.57a) $$  :
with the proviso that $f(t,r)$ admits the expansion with $r$ dependent coefficients shown above  and  :
$${\hat W}_s (Z,{\bar Z})=\sum_{N,{\bar N}}{\hat W}^{N,{\bar N}}_s Z^{-N-s/2}
{\bar Z}^{-{\bar N}-s/2}.~N=n+in/2;{\bar N}=n-in/2. \eqno (4.57b)$$
There is nothing wrong with the fact that the exponents of $Z,{\bar Z}$ in the  l.h.s of (4.57a) do not bear an $s$ dependence after premultiplying by the factor $(Z{\bar Z})^{s/2}$; the $s$ dependence is encoded in the components $ {\hat W}^{N,{\bar N}}_s$. (4.57a) is a realization of  the real-valued meromorphic field operators in terms of solutions of the continuous Toda molecue ; $\rho (r,t)$ after the change of variables is performed. It is clear from (4.57a) that the meromorphic fields are mixed and are no longer purely holomorphic nor antiholomorphic in the $Z,{\bar Z}$ variables.

Thus, 
Eqs-(4.39,4.49) 
are the equations we are  looking for.     
These are the eigenvalue equations which determines the very intricate relationship between 
$\varphi_{\lambda,{\bar \lambda}} [t(Z,{\bar Z})] $
and the weights $\Delta^\lambda_k$ ( and the antichiral ones). In order to find such a relationship one needs first to expand   :

$$\varphi_{\lambda, {\bar \lambda}}[t(Z,{\bar Z})] =
\sum_{m,{\bar m}}~A_{m{\bar m}} (\lambda, {\bar \lambda}) Z^{im} {\bar Z}^
{i{\bar m}} \eqno (4.58) $$
The reality condition on $\varphi (t)$ imposes the form of the exponents in the expansion. $(Z/{\bar Z})^{i}=e^{2t}$ is real. This fixes ${\bar m}=-m$ 
and convergence of the series implies certain restrictions on  
$A_{m,{\bar m}}$. Per example, the latter  cannot be all positive-valued functions if convergence occurs for  $m\rightarrow \infty$  or they must be in decreasing sequence. An expansion in terms of trigonometric functions of $t$ is also valid;  whereas   
 expanding in powers of $t$ is equivalent to expanding in powers of ${1\over 2}ln (Z/{\bar Z})^i$ which, in turn, can be Taylor expanded 
in powers of   :
$$ln (q+[(Z/{\bar Z})^i-q])=lnq+ln (1+[{(Z/{\bar Z})^i\over q}-1])=
lnq +[{(Z/{\bar Z})^i\over q}-1]-{1\over 2}[{(Z/{\bar Z})^i\over q}-1]^2 +....
\eqno (4.59)$$
where $q$ is chosen to ensure convergence of the Taylor series; i.e. the $ln (1+x)$ converges for $|x|<1.$ The variable $t$ is an ``angle'' variable  ranging from $[0,2\pi]$. The coefficients,  $A_{m{\bar m}}$, are  functions ( to be determined )  of the $\lambda, {\bar \lambda}$ parameters
characterizing the representations (like the weights). This is precisely where a regularization of the expectation values of the integrals 
(4.39,4.49) 
is required in the same vain that the expectation $<L_0>=a$ required a zeta function regularization for the string. To define the $|\rho (in)>$ state requires the evaluation of the $r=-\infty$ limit of the quantity  $r(-\varphi (t)$ which appears in the asymptotic regime in the solutions like 
(3.1) 
and, as specified in 
(4.21), 
requires a change in sign to assure convergence at $r=-\infty$. To define the ``in'' state requires to evaluate the limit :

$$lim_{Z,{\bar Z}\rightarrow 0}~ln(Z{\bar Z})
\sum_{m,{\bar m}}~A_{m{\bar m}} (\lambda, {\bar \lambda}) Z^{im} {\bar Z}^
{i{\bar m}} \eqno (4.60) $$
The powers $(Z/{\bar Z})^{im}$ are not well defined ( although finite, $e^{2mt}$) in the $Z,{\bar Z}\rightarrow 0$
limit. the regularization prescription must be for all values of $m$  :
$$ lim_{Z,{\bar Z}\rightarrow 0}~ln(Z{\bar Z})
A_{m,- m} (\lambda, {\bar \lambda})Z^{im}{\bar Z}^{-im}= 
A^{reg}_{m,-m} (\lambda, {\bar \lambda}) e^{2mt}.\eqno (4.61)$$
The presence of $t$ in 
(4.61) 
is  due to the ambiguity of the zero limit  and is important. From 
eqs-(4.25,4.29) 
one learns that  the ``in'' state, $|\rho_{\lambda,{\bar \lambda}}>$   bears an intrinsic $t$ dependence encoded in the function $\varphi (t)$. After computing the inner products 
; $ <\rho|\rho>=1$ this $t$ dependence drops out appearing only on the limits 
$t=0,t=2\pi$ as shown explicitly in 
(4.29). 
In this fashion the ambiguity is removed.   Therefore, choosing the infinite number of  ``coefficients'' to absorb the logarithmic singularity regularizes  the expectation values of the integrals . Per example, after the contour integrations absorb the singularities in the OPE of the Toda exponentials, the regularized expectation value of the zero modes of the $W_2(Z,{\bar Z})$ operator/generator is  :

$$<\rho_{\lambda,{\bar \lambda}}|{\hat W}_2^{N=0,{\bar N}=0}|\rho_{\lambda,{\bar \lambda}}>_{reg} ={\cal F}_2[A^{reg}_{m,{\bar m}} (\lambda,{\bar \lambda })]={\cal P}[  \Delta^\lambda_1, {\bar \Delta}^{\bar \lambda}_1 ]. \eqno (4.62)$$
and similarily for $s=3,4,5,......\infty$
$$<\rho_{\lambda,{\bar \lambda}}|W^{N=0,{\bar N}=0}_s|\rho_{\lambda,{\bar \lambda}}>_{reg} ={\cal F}_s[A^{reg}_{m,{\bar m}}(\lambda, {\bar \lambda})]=  
{\cal P} [\Delta^\lambda_k ,{\bar \Delta}^{\bar \lambda}_k.] \eqno (4.63)$$
The r.h.s of 
(4.62,4.63) 
is given by the prescription in 
(4.26b)).
The infinite number of eqs 
(4.62,4.63) 
are the ones obtained after the evaluation of 
(4.39,4.49) 
using the regularization prescription 
(4.61) for the states as well as the point-splitting for the operators involved in the definition of the $W_s$ generators.
The ${\cal F}_s;$ for $s=2,3,4.....\infty$ are an infinite number of  known functionals of the regularized ``coefficients `` $A^{reg}_{m,{\bar m}}$ that bear the sought-after $\lambda,{\bar \lambda}$ dependence encoded in the $\varphi (t)$ function given by 
(4.58)
Having an infinite family of functions in 
$\lambda, {\bar \lambda},~{\tilde \Delta}^\lambda_k,...~k=1,2......$, the integral equations 
(4.62,4.63) 
for $s=2,3,4........$,
will be sufficient ( in principle ) to specify  $A^{reg}_{m{\bar m}}(\lambda, {\bar \lambda});~m=0,1,2......$ enabling to establish the $|\lambda,{\bar \lambda}>\rightarrow |\rho_{\lambda{\bar \lambda}}>$ correspondence. Therefore,  
these  infinite number of equations, would allow us  to construct 
the states $|\rho_{\lambda {\bar \lambda}}>$ and quantize the theory exactly.  
\smallskip
\centerline{\bf 4.4 CASIMIRS}
\smallskip

One way to obtain explicit answers without having to compute the OPE of Toda 
exponentials is by recurring to the construction of the infinite number of Casimirs. In ${\bf V}$ other ways to obtain exact answers are found. The 
infinite number of involutive  $regularized$ quantum integrals of motion, the Casimirs, are    [4] :

$$I_n [(<\rho|{\hat \varphi}_{\lambda,{\bar \lambda}} (t)|\rho>_{reg})] =<\rho|{\hat I_n}|\rho>_{reg} \sim \int^{2\pi}_0~dt~(\int^t~dt' 
\varphi_{\lambda,{\bar \lambda}} (t'))^n.
 \eqno (4.64)$$

Having an explicit solution for $\varphi_{\lambda,{\bar \lambda}}(t)$ automatically yields the expectation values of the infinite number of Casimirs of the $U_{\infty}$ algebra. The explicit expression relating the infinite number of involutive conserved charges in terms of the generators of the chiral $W_{\infty}$ algebra has been given by Wu and Yu [33] :

$${\hat Q}_2=\oint~{\hat W}_2 dz;~{\hat Q}_3 =\oint~{\hat W}_3 dz.~
{\hat Q}_4 =\oint~({\hat W}_4-{\hat W}_2.{\hat W}_2)(z)dz.$$

$${\hat Q}_5 =\oint~({\hat W}_5-6 {\hat W}_2.{\hat W}_3)(z)dz;
{\hat Q}_6 =\oint~({\hat W}_6  -12 {\hat W}_2.{\hat W}_4 -12 {\hat W}_3 {\hat W}_3 +8 {\hat W}_2 {\hat W}_2{\hat W}_2
                            )(z)dz;....\eqno (4.65)$$

Similar expressions hold for the antichiral algebra. In the dimensionally-reduced $U_{\infty}$ algebra case, these expressions hold as well, where now the integrals to use are those of the type outlined in 
eqs-(4.24) 
which  replace the contour integrals in the complex plane $z,{\bar z}$ by integrals w.r.t. the $t$ variable.  The $f(t,r=-\infty)$ terms  can be set to a constant ( the zero mode is selected ). Upon evaluation of expectation values and 
a regularization  one will have then expressions of the type :
$$
<{\hat I}_2>_{reg}=E= <\int dt ~{\hat W}_2 [\rho (r=-\infty,t)]>_{reg};
~<{\hat I}_3>_{reg} = <\int dt ~{\hat W}_3 [\rho (-\infty,t)]>_{reg}.$$
$$ <{\hat I}_4>_{reg} =<\int dt ~({\hat W}_4-{\hat W}_2.{\hat W}_2)
[\rho(r=-\infty,t)]>_{reg}.
....\eqno (4.66)$$

The values of the expectation values in (4.66) should  agree with the Casimirs (4.64). These expressions could have been given in terms of the $Z,{\bar Z}$ as well. 
A question immediately arises :
How does one know that a removal of infinities by a suitable regularization procedure yields precisely eqs-(4.64,4.66) for the expectation values ? The answer lies in the $complete$ quantum integrabilty property of the quantum continuous Toda theory. If the theory is integrable the expectation values of the Casimirs ( after regularization effects) should agree with the first term of (4.64). 
If one knows $a~ priori$ the explicit $\lambda, {\bar \lambda}$ dependence of the Casimirs, a dimensional reduction yields the Casimirs of the $U_\infty$ algebra and eqs-(4.64) yield automatically the required relations sufficient to determine  the  $\lambda,{\bar \lambda }$ dependence of $\varphi (t)$ $without$ having to evaluate the OPE in 
eqs-(4.25,4.32).
 Therefore, a prior knowledge of the expectation values of the Casimir operators  of the $W_\infty$ algebras would furnish the correspondence between $|\lambda, {\bar \lambda}>\rightarrow |\varphi_{\lambda, {\bar \lambda}} (t)>$ without the need to evaluate the OPE.    

The infinite number of eqs-(4.25,4.32) 
are almost impossible to solve at the moment for the reason that we don't have the OPE rules for the exponentials of continuous Toda fields. Unless an explicit construction of the Casimirs associated with $W_{\infty},U_{\infty}$ algebras is known.  
As far as we know the Casimirs  have not been constructed. There are  special cases when we can  have exact solutions. This is discussed  in the following section.

To summarize : given a quasi-finite highest-weight irreducible representation of the $W_{\infty},{\bar W}_{\infty}$ algebras; i.e. given the generating function for the infinite number of conformal chiral ( antichiral ) weights
: ${\tilde \Delta}^\lambda (x)\Rightarrow \Delta^\lambda_k$ and the central charge ; $C $ and $b(w),\chi (characters)....$ one can ( in principle ) from 
eqs-(4.25,4.32) 
determine $\varphi (t)$
as a family of functions parametrized by $\lambda,{\bar \lambda}$.  
Since the latter are continuous parameters the energy spectrum 
(3.9,4.15)
is
$continuous$ in general. One has a continuum of highest weight states.  Below we will study a simple
case when one has a discrete spectrum characterized by the positive integers $n\ge 0$.    
Once $\varphi_{\lambda{\bar \lambda}} (t)$ is determined  the  regularized Hamiltonian operator  obeys the  equation :

$${\hat H} ~|\varphi_{\lambda {\bar \lambda}} > = E [\varphi_{\lambda{\bar \lambda}} (t)] ~ |\varphi_{\lambda {\bar \lambda}} >.       
\eqno (4.67)$$
where the Energy eigenvalue is (3.19). An explicit knowledge of 
$\varphi (t)$ can be given once 
eqs-(4.25,4.32) 
are solved.
Instead of having quantum numbers say  $|n,l,m_l...>$ like  in the hydrogen atom, per example, here  one has for ``eigenvalues'' integrals of suitable functions of $t$ which are the regularized expectation values 
of expressions involving the operator ${\hat \rho }(r,t)$. To know $\varphi (t)$ requires finding out  the values of the infinite number of coefficients, $A_n,B_n$ in 
(4.31) 
that play the role of the  ``spectra''. This is a very difficult task.  We proceed in ${\bf V}$ to find particular exact solutions.

\centerline{\bf V. Discrete Spectrum}

Below we will study a simple
case when one has a discrete spectrum characterized by the positive integers $n\ge 0$; i.e. the $|\lambda>=|n>$. We shall restore now the coupling
$\beta^2<0$ given in (2.13) . A simple fact which allows for the possibility of discrete energy states is to use the
analogy of the Bohr-Sommerfield quantization condition for periodic system. It occurs  if one opts to choose for the quantity 
$exp[\beta \varphi (t) r]\equiv exp[i\Omega r]$ which appears in (2.14); $\Omega$ is the frecuency parameter ( a constant ). For this new choice of $\varphi (t)$ the expression (3.9) for the classical energy needs to be modified in general. Below we will show that if $d(t)$ is chosen to be zero then (3.9) is still valid. When 
the dynamical system is periodic in the variable $r$ 
with periods $2\pi /\Omega$, a way to quantize the values of $\Omega$ in units of $n$ is to recur to the Bohr-Sommerfield
quantization condition for a periodic orbit :
$$ J=\oint~pdq =n\hbar       \eqno (5.1)$$ 

which reflects the fact that upon emission of a quanta of energy $\hbar \Omega$ the change in the enery level as a function of
$n$ is  [34]  :

$$  \partial E/\partial n =\hbar \Omega ={2\over 3} (2\pi)^3 (\hbar )^2\Omega \partial \Omega /\partial n.\Rightarrow $$

$$\Omega (n) ={3\over 2 (2\pi)^3 }{ n\over \hbar}. \eqno (5.2)$$
Hence the energy is 

$$E={3\over 4}(2\pi)^{-3} n^2.    \eqno (5.3)$$
which is reminiscent of the rotational energy levels $E\sim h^2l(l+1)$ of a rotor in terms of the angular momentum quantum
numbers $l=0,1,....$. In order to have a proper match of dimensions it is required to insert the membrane tension as it happens
with the string. In order to classify the physical set of states we have to have at our disposal of all of the unitary 
highest weight irrepresentations. As far as we know these have not been constructed for $W_{\infty}$. For $W_{1+\infty}$ these
have been constructed by Kac and Radul [7] and by the group [6]. These representations turned out to have a crucial importance 
in the classification of some of the Quantum Hall-Fluid states [11].

Saveliev [4] chose the $\varphi (t)$ in (3.1) to be negative real functions to assure that the
potential term in the Hamiltonian vanished at $r\rightarrow \infty$ and arrived at (3.9). 
In  case that the functions $\varphi (t)$ are no longer $<0$; i.e when $\beta \varphi r$ is no longer a real valued quantity
$<0$,  the asymptotic formula (3.9) will no longer hold and one will be forced to perform the very complicated integral !

$${\cal H} = \int~dt[-{1\over 2}\beta^2(\partial p/\partial t)^2 +({m^2\over \beta^2}) exp~[\beta\partial^2x/\partial t^2 ]
]. \eqno (5.4)$$

where $p= \beta \partial x/\partial r$ is the generalized momentum corresponding to $\rho \equiv \beta\partial^2x/\partial
t^2,$ and 
 $\mu^2 \equiv ({m^2\over \beta^2})$ is the perturbation theory expansion parameter discussed in [5]. Without loss of
generality it can be set to one. Nevertheless, eqs-(5.1,5.2) are still valid. One just needs to evaluate the
Hamiltonian at $\Omega r=2\pi p$ where $p$ is a very large integer $p\rightarrow \infty$ and take $d(t)=0$ in (3.2,3.3) ( the logarithm is illdefined, nevertheless the energy is still finite ) :

$$exp[\partial^2 x/\partial t^2]\rightarrow d(t)exp[i2\pi p] =0.~(\partial p/\partial t)^2\rightarrow (\int \varphi
dt')^2...\eqno (5.5)$$   recovering (3.9) once again.  

Are there zero energy solutions ?. If one naively set $\varphi (t) \equiv 0$ in (3.2) or set $n=0$ in (5.2) one would get a zero
classical energy. However eqs-(3.2,3.3) for the most part  will be singular and this would be unacceptable. One way  zero
energy states could be obtained is by choosing $\varphi (t), d(t)$ appropriately so that (5.4) is zero. Since one has one
equation and two functions to vary pressumably there should be an infinite number of solutions of zero energy. In the quantum case one has an extra function
to deal with $w(t)$ so it is possible to set $\varphi $ to zero and use $d(t),w(t)$ appropriately to avoid singularities. 

At first sight there could be an infinity of quantum ground states. In the discrete spectrum case the lowest of the ground states ( the lowest in energy of 
the highest weight states) corresponds to $n=0$. From this state one then builds a representation by erecting the tower of states (4.5). 
As mentioned earlier we do not know whether the discrete states are the physical ones. The no-ghost theorem has yet to be constructed.
For this reason it is of paramount importance to have a list of all unitary highest weight irreps in order to avoid negative norm
states. In the string picture one has that the central charge must be $26$ and the Regge intercept must be $a=1$. The membrane, as a 
non-critical $W_{\infty}$ string, is comprised of an infinite set of Virasoro type of strings with unusual central charges and intercepts
[25]. Therefore, we expect to have a selected value for the conformal weights $\Delta_k$ as well as the central charge.

Solving (4.67) is analogous to solving a time independent  ``Schroedinger''-like equation. Concentrating on the case that $\varphi (t) <0 $;
 the wave functional is defined :
$\Psi [\rho,t]\equiv <\rho|\Psi>$ where the state $|\rho>_{\varphi}$ has an explicit dependence on $\varphi$ which also depends
on $\lambda$ as shown in section ${\bf IV}$. Upon replacing 

$\partial p/\partial t \rightarrow -i\hbar (\partial/\partial t \delta/\delta \rho)$
as an operator acting on the $\Psi$, the time-independent equation for the wave functional becomes :

$$\int^{2\pi}_0 dt~[(-i\hbar \partial/\partial t \delta/\delta \rho)^2 +exp~\rho ] \Psi [\rho (r't');t] = 
\int^{2\pi}_0 dt~(\int^t dt'\varphi (t'))^2 \Psi [\rho (r',t');t].
\eqno (5.6)  $$

One could have written (5.6) in the momentum representation :$ \rho \rightarrow -i\hbar \delta/\delta p$ acting on the "Fourier" transfom
of $\Psi$.

The action functional is :

$$\int dt\int~{\cal D}\rho dr \Psi^+(i\hbar {\partial \over \partial r} -{\cal H})\Psi [\rho (r',t');r,t]. \eqno (5.7)$$

${\cal D}\rho$ is the functional integration measure; $r$ is the variable linked with the physical time
and the on-shell condition is just (5.6).
This is the second-quantization of the physical quantities. $\rho (r,t)$ has already been first-quantized in (3.6,3.7). 
One must  $not$ interpret $\Psi$ as a probabilty amplitude but as a field operator  which creates a
continuous Toda field in a given quantum state $|\rho>_{\varphi}$ associated with the classical configuration configuration given
by eq-(3.1) in terms of $\varphi_{\lambda} (t)$. The functional differential equation (5.6) is extremely complicated.  
A naive zeroth-order simplification will be given shortly. This is  because the $\Psi$ can have the form $\Psi =\Psi [\rho,\rho_{t'},\rho_{t't'},.......]$.                                                                  
The  equation in the momentum representation does not have that complexity but it has an exponential functional differential operator.
Evenfurther, the $\Psi$ is a non-local object. This is of no concern : In string field theory the string field is a multilocal object that depends on 
all of the infinite points along the string.

One can expand $|\Psi>$ in an infinite dimensional basis spanned by the Verma module  (4.5) associated with the state
$|\lambda>$ and let $\lambda$ run as well over all the highest weights. Given a vector $v_\lambda \epsilon  {\cal V_\lambda}$ ( Verma module)  one has :

$$|\Psi > = \sum_{\lambda}\sum_{v_\lambda}~<v_\lambda ||\Psi > |v_\lambda >.~v_\lambda \epsilon {\cal V_\lambda}    \eqno (5.8)$$

This is very reminiscent of the string-field $\Phi [X (\sigma )] =<X||\Phi (x_o)>$ where $x_o$ is the center of mass coordinate
of the string and the state $|\Phi (x_o)>$ is comprised of an infinite array of point fields :

$$|\Phi (x_o)> =\phi (x_o) |0> +A_\mu (x_o) a^{\mu +}_1 |0> +g_{\mu\nu}a^{\mu +}_1 a^{\nu +}_1 |0> +.....\eqno (5.9)$$

where the first field is the tachyon, the second is the massless Maxwell, the third is the massive graviton....In the  string
case one does not customary expand over the towers of the Verma module since these states  have zero  norm. However one should include
all of the states. The oscillators play the role of ladder-like operators acting on the ''vacuum''$|0>$ in the same manner that the Verma module is
generated from the highest weight state $|\lambda>$ by succesive application of a string of $W(z^{-n}D^k)$ operators acting on
$|\lambda>$.
The state $|\rho (r,t)>_{\varphi}$ 
It is the relative of the string state $|X(\sigma_1,\sigma_2)>$ whereas $|\Psi >$ is the relative of the string field state 
$|\Phi >$. The naive zeroth order aproximation  of the ``Schroedinger''-like equation is of the form :

$$[\partial_t^2 \partial_y^2 +e^y] \Psi (y,t) =E\Psi (y,t). \eqno. (5.10)$$

A change of variables :$x=2e^{y/2}$ converts (5.10) into  Bessel's equation after one sets $\Psi (y,t) =e^t\Phi (y)$ or
equal to $e^{-t}\Phi (y)$  :   

$$  (x^2\partial^2_{x^2} +x\partial_x +x^2-4E)\Phi (x)=0. \eqno (5.11)$$

and whose solution is :$\Phi (x) =c_1{\cal J}_\nu (x) +c_2 {\cal J}_{-\nu} (x)$
where $\nu \equiv 2\sqrt E$ and $c_1,c_2$ constants.

The wavefunctional is then a linear combination of  :

$$\Psi [\rho(r',t');t] = e^t\int\int~dr'dt'[c_1{\cal J}_\nu (2e^{\rho (r',t')/2}) +c_2{\cal J}_{-\nu} (2e^{\rho (r',t')/2})]  \eqno (5.12)$$
or the other solution involving $e^{-t}...$.

One may notice that  discrete energy level (in suitable units such as  $\nu=2\sqrt E=n$) solutions are possible. Earlier
we saw in (5.2) that $E(n)\sim n^2$ so $\sqrt E \sim n$. Therefore setting $2\alpha e^{y/2\alpha}=x$ where $\alpha$ is a 
suitable constant allows to readjust $\nu =2{\sqrt {\alpha E}}=n$. The
Bessel functions will have nodes at very specific points . The solutions in this case will be given in terms of ${\cal J}_n$ and
the modified Bessel function of the second kind, $K_n$. These solutions are tightly connected with the boundary conditions of
the wave-functional.

The other instance in which exact solutions can be obtained is when the OPE of the Toda exponentials is available. There exists a particular class of solutions to the $3D$ continuos Toda field equation that bears a direct relationship to the Liouville equations :
$${\partial^2 \over \partial z \partial {\bar z} }u (z,{\bar z},t)={\partial^2 e^u \over  \partial t^2}. \eqno (5.13)$$

The ansatz [4] 

$$e^{u(z,{\bar z},t)} =(\alpha t^2 +\beta t+\gamma)e^{\phi (z,{\bar z})}\Rightarrow 
~\partial_z \partial_{\bar z} \phi (z,{\bar z})=e^{\phi (z,{\bar z})}             \eqno (5.14)         $$
so a particular class of solutions of (5.13) in terms of those solutions of the Liouville equation (5.14) can be  found . The solutions of (5.13) are tightly connected to 
Killing symmetry reductions of Plebanski heavenly equations for $4D$ self dual gravitational  backgrounds. Evenfurther, these hyperkahler/heavenly  backgrounds exhibit duality symmetries via 
the Legendre transform. For a recent discussion on the status of   
non-abelian  duality symmetries associated with  the non-linear supersymmetric $\sigma$ models and $WZNW$ models relevant to  superstring theory, see 
Sfetsos [42].  
In [14] we have recently  derived a straightforward  origin of the analog of $S,T$ duality symmetries for all $p$ branes using  a reformulation of extended objects as composite antisymmetric tensor field theories of the volume-preserving diffeomorphism group  based on the work of [13]. The upshot of this formulation is that $S,T$ duality emerge in a very natural way from the start. The quantization program remains yet to be done. 

Thus, for the special class of solutions (5.13) one can compute the OPE of the 
Toda exponentials in terms of Liouville field theory  following the 
fusion rules constructed by Bilal,Gervais and Schnittger [32]. Quantum $tau$ functions can also be explicitly built [40].   
For this particular case we can have also  explicit results of  ${\bf IV}$.

\centerline{\bf VI. CONCLUSION AND CONCLUDING REMARKS}

An exact integrable set of quantized solutions of the spherical membrane moving in flat target backgrounds has been obtained. Such  a special class of solutions are those related to dimensional reductions of $SU(\infty)$ YM theories from ten to four 
dimensions. The latter are constructed in terms of instanton solutions in $D=4$ that can be related to the continuous Toda theory after an ansatz is made. Not surprisingly, the system is integrable. We don't know the fate of more general class of solutions of the YM equations nor what happens for surfaces whose topology is not spherical nor the case of membranes moving in curved backgrounds. This deserves further study. We hope to have advanced the need to built unitary highest weights irreps of $W_\infty$ algebras and the construction of the OPE of continuous Toda exponentials required to furnish the full non-perturbative membrane's spectrum.  

What will happen with other $p-branes/extendonds$ ? A reformulation of $p$ branes has been recently been  constructed by the author [14] based on [13]. Extended objects can be seen as new $composite$ antisymmetric tensor field theories of the volume-preserving diffs group. 
The positive side to this reformulation is not only that the volume-preserving diffs symmetry is  manifest but $S,T$ duality emerge in a very natural way from the very beginning. The next step is to find explicit classical solutions 
and use the representation theory of the volume-preserving-diffs to construct the spectrum. 
Membranes appear in a wide variety of physical models .

1-.It has been argued that a four dimensional 
anti-deSitter spacetime, $AdS_4$, whose boundary is  $S^{2}\times S^1$, could be realized as a membrane at the end of the
universe. In particular, singleton field theory can be described on the boundary of $AdS_4$ where singletons are the most
fundamental representations of the de Sitter groups [36]. Moreover, on purely kinematical grounds, infinitely many massless
states of all spins (massless in the anti-de-Sitter sense) can be constructed out of just two singletons ( preons). In
particular, the $d=4~N=8$ Supersingleton field theory formulated on the boundary of $AdS_4$ bears a connection with the
supermembrane moving on  $AdS_4\times S^7$ . The rigid $OSp(8|4)$ symmetry acts as the superconformal group on the boundary 
$S^{2}\times S^1$ [36].  In view of this it is important to study if there is any connection between the wave-functional
behaviour of ${\bf V}$ at the boundaries and singleton field theory. The supersymmetric Toda equation has been discussed by [5], thus in principle the results found here could be extended to the supersymmetric case.

Roughly speaking, a membrane is comprised of an infinite number of strings. Thus the membrane can be seen as a coherent state of
an infinite number of strings . This is reminiscent of the Sine-Gordon soliton being the fundamental fermion of the massive
Thirring model, a quantum lump [34]. The lowest fermion-antifermion bound state (soliton-antisoliton doublet) is the fundamental
meson of Sine-Gordon theory. Higher level states are built from excitations of the former in the same way that infinitely many
massless states can be built from just two singletons.

2-.Edge states : Recently [11], the set of unitary highest weight irreps of $W_{1+\infty}$ have been used to algebraically characterize the low 
energy edge-excitations of the incompressible ( area preserving) Quantum Hall Fluids. Experimental data matching the Jain hierarchy
for certaing filling factors was identified with certain minimal models. A surprising feature was found : Non Abelian structures
were discovered and neutral quark-like excitations with $SU(m)$ quantum numbers and fractional statistics as well. This merits
a further investigation from the membrane picture : a dimensionally reduced $SU(\infty)$ Yang-Mills theory.                        

3-.Noninteracting multi string solutions in curved spacetimes were studied in 
[37]. A single world sheet simultaneously 
decsribed many different and independent strings with no analog in flat space which appears as a consequence of the coupling
of the strings with the curved spacetime (de Sitter). This ``multistring'' picture fits with the membrane picture although one has to look at the propagation in 
anti de Sitter backgrounds.

4-.Perhaps the most relevant physical applications of the membrane quantization program will be in the behaviour of black hole
horizons [8]. These have been described in terms of a dynamical surface whose quantum dynamics is precisely that of a
relativistic membrane. Thermodynamical properties like the entropy and temperature of the black hole were derived in agreement
with the standard results. Results for the level structure of black holes were given. A ''principal'' series of levels was found
corresponding to the quantization of the area of the horizon in units of the ''area quantum'' :$A=nA_o.~A_o =8\pi$. From each
level of this principal series starts a quasi-continuum of levels due to the membrane's excitations. The connection between black hole physics and non-abelian Toda theory has been studied in [9]. $W$ gravity was formulated as chiral embeddings of a Rieman surface into $CP^n$. Toda theory plays a crucial role as well [10].

5-. The ordinary bosonic string has been found to be a special vacua of the $N=1 $ superstring [12]. It
appears that there is a whole hierarchy of string theories : $w_2$ string is a particular vacua of the $w_3$ string and so
forth......If this is indeed correct one has then that the (super) membrane, viewed  as  noncritical $W_{\infty}$ string theory,
is, in this sense, the universal space of string theory. 
The fact, advocated by many, that a Higgs symmetry-breakdown-mechanism of the infinite number
of massless states of the membrane generates the infinite massive string spectrum fits within this description.

6. Finally, we hope that the essential role that Self Dual $SU(\infty)$ Yang-Mills theory has played in the origins of the
membrane-Toda theory, will shed more light into the origin of duality in string theory [38,39]. For a review of duality in string theory and the status of string solitons see [40]. An important review of extended conformal field theories
 see [43].

As of now  we must have all unitary irreps of $W_{\infty}$ and be able to construct the OPE of the Toda exponentials to  fully exploit the results of ${\bf IV}$. The discrete spectrum solution warrants a further investigation and the supersymmetric sector as well.

\smallskip

ACKNOWLEDGEMENTS. We thank M.V. Saveliev for many helpful suggestions concerning the exact quantization program of the continuos Toda theory. 
To G.Sudarshan, Y.Ne'eman, C.Ordonez, J.Pecina, B. Murray for discussions  and to the Towne family for their kind and
warm hospitality in Austin, Texas. This work was supported in part by a ICSC, World Laboratory Fellowhip.

 \smallskip

\centerline {\bf REFERENCES}

1. C. Castro : Journal  of Chaos, Solitons and Fractals. Special issue 

on ``Quantum Mechanics in Rigged Space-Times `` Elsevier Science Publishers 

April (1996) 439-453.

2-.A. Ivanova, A.D. Popov : Jour. Math. Phys. {\bf 34} (1993) 674. 

3. J. Hoppe : "Quantum Theory of a Relativistic Surface" MIT Ph.D thesis (1982)

4-M.V. Saveliev : Theor. Math. Physics {\bf vol. 92}. no.3 (1992) 457.

5.A.N.Leznov, M.V.Saveliev, I.A.Fedoseev : Sov. J. Part. Nucl. {\bf 16} no.1 (1

985) 81.

A.N.Leznov, M.V.Saveliev : "Group Theoretical Methods for Integration of Nonlin

ear Dynamical Systems " Nauka, Moscow, 1985.

6. H.Awata, M.Fukuma, Y.Matsuo, S.Odake :"Representation Theory of the 

$W_{1+ {\infty}}$ Algebra".  RIMS-990 Kyoto preprint , Aug.1994.

S.Odake : Int.J.Mod.Phys. {\bf A7} no.25 (12) 6339.

7. V.Kac, A. Radul : Comm. Math. Phys. {\bf 157} (1993) 429.

8. M. Maggiore :''Black Holes as Quantum Membranes : A Path Integral Approach '' 

hepth-lanl-9404172. 

9.J.L. Gervais, M.V. Saveliev : Nucl. Phys. {\bf B 453} (1995) 449.

10. J.L. Gervais, Y. Matsuo : Comm. Mtah. Phys. {\bf 152} (1993) 317.

11. A. Capelli, C.A.Trugenberger, G.R. Zemba :" Quantum Hall Fluids as 

$W_{1+\infty}$ Minimal Models"DFTT-9/95

12. N. Berkovits, C.Vafa :Mod. Phys. Letters {\bf A9} (1994) 653.

13. E. Guendelman, E. Nissimov, S. Pacheva . `` Volume-preserving diffeomorphis

ms versus Local Gauge Symmetry `` hep-th /950512 

A.Aurilia, A. Smailagic, E. Spallucci : Phys. Rev. {\bf D 47} (1993) 2536.

14. C. Castro : ``p-branes as Composite Antisymmetric Tensor Field Theories ``

hep-th/9603117.

15. D.B. Fairlie, J. Govaerts, A. Morozov : Nucl. Phys {\bf B 373} (1992) 214.

D.B. Fairlie, J. Govaerts : Phys. Lett {\bf B 281} (1992) 49.             .

16-. B. Biran, E.G.F. Floratos, G.K. Saviddy : Phys. Lett {\bf B 198} (1987) 32

9.

17-.E.G.F. Floratos, G.K. Leontaris : Phys. Lett {\bf B 223} (1989) 153

18-. M. Toda : Phys. Reports {\bf 18} (1975) 1.

19. R. Zaikov : Phys. Letters {\bf B 211} (1988) 281.

Phys. Letters {\bf B 266} (1991) 303.

20. M.Duff : Class. Quant. Grav. {\bf 5} (1988) 189.

21. Y. Ne'eman, E. Eizenberg : `` Membranes and Other Extendons `` World 

Scientific Lecture Notes in Physics. {\bf vol. 39} (1995).

22. U. Marquard, R. Kaiser, M. Scholl : Phys. Lett {\bf B 227} (1989) 234.

U. Marquard, M. Scholl : Phys. Lett {\bf B 227} (1989) 227.

23 .J. de Boer : "Extended Conformal Symmetry in Non-Critical String Theory" . 

Doctoral Thesis. University of Utrecht, Holland. (1993).

24. J.de Boer, J. Goeree : Nucl. Phys. {\bf B 381} (1992) 329.

25.H.Lu, C.N. Pope, X.J. Wang: Int. J. Mod. Phys. Lett. {\bf A9} (1994) 1527. 

H.Lu, C.N. Pope, X.J. Wang, S.C. Zhao : "Critical and Non-Critical $W_{2,4}$ 

strings". CTP-TAMU-70-93. hepth-lanl-9311084.

H.Lu, C.N. Pope, K. Thielemans, X.J. Wang, S.C. Zhao : "Quantising Higher-Spin 

String Theories "  CTP-TAMU-24-94. hepth-lanl-9410005.

26. E. Bergshoeff, H. Boonstra, S. Panda, M. de Roo : Nucl. Phys. {\bf B 411} (1

994)  717.

27.R.Blumenhagen, W. Eholzer, A. Honecker, K. Hornfeck, R. Hubel :" Unifying 

$W$ algebras".   Bonn-TH-94-01 April-94. hepth-lanl-9404113. 

Phys. Letts. {\bf B 332} ( 1994) 51

28 . M.A.C. Kneippe, D.I. Olive : Nucl.Phys. {\bf B 408} (1993) 565. 
  
Cambridge Univ. Press. (1989). Chapter 6,page 239.

29 .C.Pope, L.Romans, X. Shen : Phys. Lett. {\bf B 236} (1990)173. 

30. I.Bakas, B.Khesin, E.Kiritsis : Comm. Math. Phys. {\bf 151} (1993) 233.

D. Fairlie, J. Nuyts : Comm. Math. Phys. {\bf 134} (1990) 413.

31 . C.Castro :Phys. Lett {\bf 353 B} (1995) 201.

32. J.L. Gervais, J. Schnittger : Nucl. Phys {\bf B 431} (1994) 273.

 Nucl. Phys {\bf B 413} (1994) 277.

33.S. Coleman : "Aspects of Symmetry " Selected Erice Lectures.  

34. F. Yu, Y.S. Wu : Journal. Math. Phys {\bf 34} (1993) 5851-5895

.

35. B. de Wit, J. Hoppe, H. Nicolai : Nucl. Phys. {\bf  B 305} (1988) 545.

B. de Wit, M. Luscher, H. Nicolai : Nucl. Phys. {\bf  B 320} (1989) 135.

36.E.Bergshoeff, A.Salam, E.Sezgin ,Y.Tanii : Nucl. Phys. {\bf B 305} (1988) 

497.

E.Bergshoeff, E.Sezgin ,P.Townsend  :  Phys. Letts. {\bf B 189} (1987) 75.

37 . N.Sanchez, H.de Vega : "String Theory in Cosmological Spacetimes" 

LPTHE-Paris-95-14

38. M.Duff, R. Minasan :''Putting String/String Duality to the Test'' 

CTP-TAMU-16/94.  hepth-lanl-9406198.

39. E.Witten : "String Theory Dynamics in Diverse Dimensions" IASSNS-HEP-95-18.

hepth/9503124

40. M. Duff, R. Khuri, J.X. Lu : Phys. Reports {\bf 259} (1995) 213-326

41. S. Kharchev, A. Mironov, A. Morozov : Theor. Math. Phys. {\bf 104} (1995)

866.

42. K. Sfetsos : ``Nonabelian Duality, Parafermions and Supersymmetry `` hep-th

/9602179. 

43. P. Bouwnegt, K. Schouetens : Phys. Reports ${\bf 223}$ (1993) 183.

\bye